\documentclass[12pt,preprint]{aastex}
\slugcomment{Accepted  for publication  
in {\it the Astrophysical Journal}} 
\def\lax {\ifmmode{_<\atop^{\sim}}\else{${_<\atop^{\sim}}$}\fi}  
\def\gax {\ifmmode{_>\atop^{\sim}}\else{${_>\atop^{\sim}}$}\fi}  
\def\gtorder{\mathrel{\raise.3ex\hbox{$>$}\mkern-14mu
             \lower0.6ex\hbox{$\sim$}}}
\def\etal { et al. }
\def\sax{{\em BeppoSAX\/}}

\begin{document}

\title{\sax\ Observations of the Power and Energy Spectral 
Evolution in {\em the} Black Hole Candidate XTE J1650-500}

\author{Enrico Montanari\altaffilmark{1}, Lev Titarchuk\altaffilmark{1, 2} 
and Filippo Frontera\altaffilmark{3} }

\altaffiltext{1}{Dipartimento di Fisica, Universit\`a di Ferrara, Via 
Saragat 1, I-44100 Ferrara  and IIS Calvi, 
Finale Emilia (MO), Italy; montana@fe.infn.it} 
\altaffiltext{2}{George Mason University/Center for Earth
Observing and Space Research, Fairfax, VA 22030; US Naval Research
Laboratory, Code 7655, Washington, DC 20375-5352;   
Goddard Space Flight Center, NASA,  code 663, Greenbelt  
MD 20771 USA; Dipartimento di Fisica, Universit\`a 
di Ferrara, Via Saragat 1, I-44100 Ferrara, Italy; lev.titarchuk@nrl.navy.mil}
 \altaffiltext{3}{Dipartimento di Fisica, Universit\`a di Ferrara, Via 
Saragat 1, I-44100 Ferrara, Italy and 
INAF/IASF, Bologna, Via Gobetti 101, I-40129, Bologna, Italy;  frontera@fe.infn.it;} 

\begin{abstract}
We study the time variability and spectral evolution of the Black Hole 
Candidate source XTE J1650-500 using the \sax\ wide energy range (0.12-200 keV) 
observations performed during the 2001 X--ray outburst. The source evolves 
from a low/hard state (LHS) toward a high/soft state (HSS). 
In all states the emergent photon spectrum  is described  by the  sum of 
Comptonization  and soft (disk) blackbody components. In the LHS, the 
Comptonization component dominates in the resulting spectrum.
On the other hand, during 
the HSS observed by  \sax\,
the soft (disk) component is already dominant. 
In this state the Comptonization part of the spectrum is much softer than that in the LHS
(photon index $\Gamma$ is $\sim 2.4$ in the HSS  vs. 
$\Gamma\sim 1.7$ in the LHS). 
In the  \sax\ data we find a strong signature of the index saturation with the mass accretion rate which can be considered as an observational evidence of the converging flow (black hole) in XTE J1650-500.       
We  
derive power spectra (PS) of the
source time variability in  different spectral states  as a function of energy band.
When the source undergoes a transition to softer states, the PS as a whole is 
shifted to higher frequencies which can be interpreted as 
a contraction  of the Compton cloud     during hard-soft spectral evolution.
It is worthwhile to emphasize a detection of a strong low-frequency  red noise component  in the  HSS PS which  can be considered  a signature 
of the presence of the  strong extended  disk in the HSS.
Also  as a result of our data analysis, we  find  a very weak sign of K$_{\alpha}$ line appearance in this \sax\   
data set.
This finding  does not confirm  previous claims   by Miniutti et al.   on the presence   of a broad and strongly relativistic  iron emission line in this  
particular   set of \sax\  data. 
\end{abstract}


\section{Introduction}
XTE J1650-500  was discovered by the
Rossi X-Ray Timing Explorer ({\it RXTE}) on 2001 September 5 \citep{rem01}
and subsequently reached a peak X-ray intensity
of 0.5 crab. 
Subsequent observations established that XTE J1650-500 is
a strong black hole candidate based on
its X-ray spectrum and variability in the X-ray light curve
\citep{mark01,rs01,w01,h03,k03,r04}. The radio counterpart was discovered with the
Australia Telescope Compact Array (ATCA) by \cite{groot01}. Further radio 
observations sampled the behavior of XTE J1650-500 along all its X-ray states 
\citep{corb04}. 
\cite{oro04} found the period of the binary system to be $7.63\pm 0.02$ h, and 
the mass function [low limit of black hole (BH)  mass] to be $2.7\pm0.6$ M$_\odot$.

We define {the} spectral state hardness  according to the peak position in the $EF(E)$-diagram 
related to X-ray spectrum $F(E)$ (in units of  erg keV$^{-1}$s$^{-1}$) where $E$ is {the} photon energy
[see e.g. Figure 3 in \cite{grove}]. Following this definition, we call  a low/hard   
state (LHS) {the spectral state in} which  the  maximum  of the $EF(E)-$diagram is located  
at tens of keV,
while we call  a high/soft state (HSS) the  state where the $EF(E)$-maximum is 
located  at  a few keV.
%
%
Moreover,  we call  an intermediate state (IS)  {the state in which} the source undergoes a  
spectral transition  between the low/hard and high/soft states.  
We observed XTE J1650-500  three times with \sax\ during the 2001 outburst (September 11--12, 
21--23, October 3--4, see Table~\ref{tbl-1}), during which the source evolves 
from the  LHS toward  the HSS.
%
%
The observed outburst is 
a perfect case to study a source transition  from the LHS to  the HSS. 

The long-term monitoring of Cyg X-1 has revealed not only two distinct 
energy spectra and transitions between them but it has also  established that 
the corresponding time variability power spectra (PS) evolve   along with the photon spectra
[see \cite{ST06}, hereafter ST06,  and \cite{st07}, 
hereafter ST07]. 
Recently the evolution of the power density spectra throughout the different X-ray states 
of BH X-ray binaries has been also extensively studied in many BHs by \cite{kvk08}, 
hereafter KVK08.

The persistence of the simultaneous PS and X-ray energy spectrum evolution
suggests that the underlying physical process and conditions which give 
rise to the PS are tied to the corona; and, furthermore, 
that this process varies in a well defined manner as the source progresses 
from one spectral  state to another. 

Moreover the same evolutions are seen in so many galactic X-ray binary BH sources
(see, for example, detailed discussion of these evolutions
in Cyg X-1, GRO 1655-40 and GRS 1915+ 105 in ST07)
which vary widely in both luminosity (presumably with mass accretion rate) 
and state. This  fact suggests that 
the physical conditions controlling the photon spectral  and the  PS evolutions
are similar.  
In particular, quasiperiodic oscillation (QPO) frequencies  are 
characteristics of the power spectra  for  these sources.  Given that  
{correlations between low and high frequency QPOs and between low frequency
QPOs and break frequencies in their PS have been found for a number of them}
%
%
\citep[see][]{pbk99,bpk02}, one can suggest that QPO phenomenon can be  
an universal property of  all  accreting compact systems.

In particular, the PS of black hole binaries in hard states is dominated 
by a component, which has a specific shape roughly described by a broken power-law 
({\sc bknpl}). 
The low-frequency part of the {\sc bknpl} is mostly flat, while the higher frequency
part above the ``break'' frequency $\nu_{br}$ is a power law  $\nu^{-\alpha}$ which index  $\alpha$
varies between 1 and 2. 
In fact, the   PS of BHs  are more  complicated than just a broken power law [see \cite{tsa07}, \cite{ts08}, hereafter TSA07, TS08 respectively, and KVK08 for comprehensive study of the BH power density spectra in different X-ray spectral  states].

It is also well established 
that the fractional root-mean-square (rms) variability in a source light curve decreases  while 
QPO low-frequency (and  $\nu_{br}$)  increases
as the source evolves from the LH state to the HS state [see e.g.  TS08, KVK08].
 
\citet{L97}  was the first to suggest a model for this  time variability production 
in the accretion powered X-ray sources. 
He  considered small amplitude local fluctuations in the accretion rate at each radius, 
caused by small amplitude variations in the viscosity, and then studied  the effect of 
these fluctuations on the 
accretion rate at the inner disc edge. His linear calculations show that if the 
characteristic time-scale of the viscosity 
variations  is everywhere comparable to the viscous (inflow) time-scale, and 
if the amplitude of the variations is independent of radius, then the power 
spectrum of luminosity fluctuations is a power-law of index $\alpha = 1$, namely  $\propto\nu^{-1}$.  If the amplitude of the 
variations increases with radius, the slope of the power spectrum of the luminosity variations 
$\alpha$ is steeper than 1. Lyubarskii pointed out that he had no physical model for the 
cause of such fluctuations. 
\citet{utt05} pointed out that  rms-flux relation is naturally explained in 
the framework developed by Lyubarskii.

TSA07 formulated and solved the problem of local driving 
perturbation diffusion
in a ``disk-like'' configuration (which can be either a geometric thin Keplerian 
accretion disk or a Compton cloud).  The problem of the diffusive propagation of the 
spatially distributed  high-frequency 
perturbations is formulated  as a  problem in terms of  the diffusion equation 
for  the surface density perturbations. 
This equation is combined with the appropriate boundary conditions.  
The formulation of this problem and its solution are  general and  classical. 
The emergent power spectrum  as a result of   the perturbation diffusion in a given  bounded configuration (either Compton cloud or Keplerian disk) has a white-red noise (WRN)  shape which can be approximated by a broken power-law ({\sc bknpl}) with two  indices  0 and $\alpha\gax1$ respectively. 
The parameters of the WRN  PS are a timescale of the diffusion propagation of the 
local perturbations $t_0$ and the power-law index of the viscosity distribution over radius.

Note  that the shapes of X-ray photon spectra of black hole 
(BH) and neutron star (NS)  
are generic: they consist of Comptonized blackbody-like ({\sc bb}) 
components. These Comptonization components  are seen  as  a power-law 
({\sc pl}) at  energies higher  than the {\sc bb} characteristic energies.
Sometimes one needs an exponential rollover in order to terminate the {\sc pl} component 
at high energies. The main difference is that NS spectra are usually softer for the 
same state. Another difference in these spectra is that in the NS case there are
two Comptonized BB  components (not one as in BHC spectra) which can be related to the emission from  the disk (BB color temperature about 1 keV) and NS surface (BB color temperature about
2.5 keV [see \cite{ts05}].

\cite{bmc}, hereafter TMK97, introduced the Generic Comptonization model 
({\sc bmc} model in XSPEC) that takes into account the dynamical Comptonization
(converging inflow, expected in the vicinity of the central object)
along with the thermal Comptonization.  Note that
this model can be applied to fit the observed spectra of NS and
BHC sources.

In this Paper we present the results of our evolution study of the 
energy and power spectra  in XTE J1650-500 using the \sax\ observations and  using the 
aforementioned theoretical considerations of photon and power spectral formations. 
{\it We find that  this evolution is similar to that observed 
from Cyg X-1} (see e.g. ST06 and TSA07).

 \cite{min}, hereafter M04, reported spectral results from three \sax\ 
observations of 
XTE J1650-500   during its 2001/2002 outburst. 
{By performing the analysis in the energy range from 1.5 to 60 keV 
for the TOO-1 spectrum and in the range from 1.5 to 200 keV for the other 
two spectra}, they reported the presence of a broad and strongly relativistic 
Fe K$_{\alpha}$ line with an EW of  about 200 eV, depending on TOO. 
We reanalyze the same \sax\  data set adopting the entire BeppoSAX energy 
range (0.12$-$200 keV) in order to check  
the presence  of {a strong K$_{\alpha}$ line} in the data. {We find
a very weak  sign of K$_{\alpha}$ line appearance.}

Results of our detailed spectral analysis are the subject
of our next  paper (Montanari et al. 2008 in preparation). 
Here we present results on the simultaneous evolution  of  the spectral and 
temporal properties and compare them with that 
in other black hole (BH) sources (see  ST06 and  ST07).

Our approach is two-fold: On one side we tried to fit all  photon and power 
spectra with 
the same generic simple model, in order to compare them directly. On the other hand, we study 
the source in the framework of a physical model in the attempt to find
a self-consistent scenario that  can help us  to understand the Physics  
of  the  spectral and timing evolution  that we observe.

In \S 2 we briefly describe the {\em Beppo}SAX data from XTE J1650-500.
Details of our spectral fitting  are presented  in \S 3. { We discuss an
observational evidence of a converging flow in \S 3.1.}
We present the 
results of our  analysis of K$_\alpha$ appearance in the data in \S 3.2.
Description 
of XTE J1650-500 power spectra and QPO identification are presented in \S 4 
where  we also discuss a correlation of PS   vs spectral state (photon 
index) revealed in XTE J1650-500.
Discussion and conclusions follow  in \S 5.

\section{Observations and data reduction}

XTE J1650-500 was observed on September 11--12, 21--23 and October 3--4
2001 during three Target of Opportunity
Observations (TOOs) by the Narrow Field Instruments (NFIs) on board
\sax\ \citep{boella}. Table \ref{tbl-1} reports the
observation log along with the on--source exposure times.
The NFIs include a Low--Energy
Concentrator Spectrometer \citep[LECS, 0.1--10 keV;][]{parmar},
three Medium--Energy Concentrator Spectrometer \citep[MECS, 1.5--10 keV;][]
{boellab}, a High--Pressure Gas Scintillation Proportional
Counter \citep[HPGSPC, 4--120 keV;][]{manzo}, and a Phoswich Detection
System \citep[PDS, 15--200 keV;][]{ff}. 
The SAXDAS data analysis package is used for processing data. 

For each instrument we perform the spectral and temporal analysis in the 
energy range in which the response function is well determined; given
the high statistics of the source the energy range is 0.12--4.0 keV for
the LECS, 2.5--10 keV for the MECS, 8--30 keV for the HPGSPC, 
and 15--200 keV for PDS. Note that, at  the given high count rate,  the 
available LECS response function is not reliable, therefore the 
SAX/LECS Matrix Generation Program ``lemat'' is used to obtain the 
appropriate response function of the instrument.  MECS unit 1 
was not operative.   Spectra from MECS unit 2 and 3 are merged. 
The spectra so obtained are rebinned  taking  into account the energy resolution of the 
instruments in order to get  independent data points.

Previously   M04 analyzed  the same set of  \sax\  data
 integrating  the source spectra over  the entire duration 
of  each of the three observations.
However, this integration time could provide distorted spectra, given that the source 
shows a significant spectral variability with time on shorter time scales.
This effect was first revealed  studying the time variability 
of three different hardness ratios
$R[(4-10){\rm keV}/(2-10){\rm keV}]$, $R[(10-30){\rm keV}/(2-10){\rm keV})]$, 
$R[(20-60){\rm keV}/(2-10){\rm keV}]$ with bins of 200 s  integration
time. The results, that will be reported in detail elsewhere (Montanari et al. 2008, 
in preparation), show that, depending on the epoch, significant 
(up to 30\%) spectral 
variability is present in time intervals ranging from 1 to 10 ks. 
Therefore, unlike M04, we integrated the spectra over the longest time intervals 
compatible with a negligible 
(less than 5\%)  source spectral variability. 

Namely, in the TOO-1 the spectra were divided in 12  time intervals and numerated 
from I101 to I112,  in the TOO-2 they were divided  in 19 intervals
(from I201 to I219), while in the TOO-3 they were divided in 8 intervals (from I301 to I308).
Note that LECS data are not available for five  intervals of TOO-2.

In Figure~\ref{lc} we display the source light curve related to  the period from 
September 11 to October 4, 2001  for  the 2--3 keV and 20--60 keV energy 
bands. The integration time of each bin is that of the IDs intervals  used for 
spectral analysis. 
In Figure ~\ref{lc}  one can  clearly  see 
an overall anti-correlation between the two light curves related to these two energy bands
during TOO1 and TOO-2. 
Note that the transition 
to a softer spectrum becomes prominent  in TOO-2 and continues during TOO-3 observations. 
For TOO-2 (see Fig.~\ref{lc}),  a variable rate is evident  at low energies,
unlike an almost stable count rate at high energies.
 This different time behaviour at 
different energies can be also seen  from  Fig.~\ref{rat} where we compare 
the energy behavior of MECS mean count rates in the time intervals I202 and I207.

We model the energy spectra using XSPEC 11.2.0 software package  \citep{Arnaud96}.  
 A systematic error of 1\% is added in order to take into account unavoidable uncertainties in the response functions.  
In the multi--instrument fits we leave free to vary the relative
normalization of LECS, HPGSPC and PDS with respect
to MECS and  the derived unfolded spectra are 
renormalized to the MECS level 
for a clarity of display.

For the temporal analysis we use  standard fast Fourier techniques.
Source PSs are normalized to rms units using  subtraction of the 
Poissonian noise level. Dead time corrections have been taken into account
according to \cite{klis}. The binning time for the analysis is  a  few ms.
PSs are calculated using data stretches of $\sim 100$ s and then averaged.
In this way we obtain the reliable  PSs in a frequency range from ~0.01 to ~100 Hz.
Unfortunately we cannot  provide much longer time 
stretches to extend the frequency pass band to lower values   due to 
frequent interruptions in the data rate.
We present only data points which are  statistically significant (see more 
details in \S 4). 

The uncertainties of the model parameters are given at 90\% confidence level, while
those shown in all figures are 1$\sigma$ errors.

\section{X-ray energy continuum spectra}
\label{s:spectra}
In our analysis, all derived spectra (39) cover the 0.12--200 keV 
energy band, except 5 intervals in which the LECS data are not
available. In this case the energy band was from 2.5 to 200 keV.  

To describe the continuum spectrum we use the Bulk Motion Comptonization model
({\sc bmc} in XSPEC)  \citep{bmc}
which is {\it a generic Comptonization model}. This model can be used if the 
photon energy is less than the mean electron  energy of the Compton
cloud  $E_{av}$. The choice of this particular 
theoretical model is provided by the robust nature of the BMC model
for different spectral states and independence of the  specific type of
Comptonization scenario involved. The BMC model spectrum  is the sum of
a  blackbody (BB) component (which is the disk radiation directly seen by
the  observer)
 and  the fraction of  a  BB component Comptonized in the corona.

The model has four parameters: 
the color temperature $kT_{bb}$ of a thermal photon spectrum (blackbody, BB),
the energy spectral index $\alpha=\Gamma-1$,  where $\Gamma$ 
is the {\sc pl} photon index, the parameter $A$, which  is related 
to the weight [$A/(1+A)$] of the Comptonization component, 
and  the normalization of the BB component $C_N$. The BMC model is 
valid for the general 
case of Comptonization when both bulk and thermal motion are included. 

For the thermal Comptonization, $E_{av}$ is related
to electron temperature $E_{av} \sim 2kT_e$. When the bulk motion 
Comptonization is dominant,
$E_{av}$ is related to the bulk kinetic energy of the electrons 
$E_{av}\lax m_e c^2$. 
In the thermal Comptonization case, for energies less than $E_{av}$, 
COMPTT\footnote{COMPTT model was introduced by \cite{t94}, \cite{tl95} and 
\cite{ht95} to describe the spectrum of thermal Comptonization
in the whole energy range.} and
BMC models are very similar. The thermal Comptonization and  the dynamical 
(bulk motion) Comptonization are presumably  responsible for the spectral formation
in the hard state and the soft state, respectively. Therefore, one needs 
a general model such as  the BMC that describes the spectral shape regardless of the 
specific type of Comptonization. 
Although the thermal Comptonization model can properly fit the spectral 
shape  of the observed spectra 
for all spectral states, it would give physically unreasonable values of
the best-fit parameters for the soft state spectra. 
Notably, \cite{w06}   found  optical depth $\tau$ of  the 
Compton Cloud inferred from  the COMPTT
model  that significantly decreases towards  the HSS. It is very difficult 
(in the framework of any reasonable physical model) to
justify  this tendency of $\tau$ to decrease   when the  mass accretion rate 
increases during the hard-to-soft state transition (see e.g. ST06).    
However the spectral index $\alpha$ inferred using the best-fit COMPTT 
parameters $\tau$ and $kT_e$  is a physical characteristic of the Comptonization 
process.  In fact, the  spectral index $\alpha$   is the reciprocal of the Comptonization 
parameter  $Y$ [see e.g. \cite{btk07}] and it is  independent of any type  
and model of Comptonization (thermal or bulk).

It is important to emphasize the principal difference and similarity in using 
blackbody-like spectra ({{\sc bbody} in XSPEC}) and multicolor disk spectra
({\sc diskbb in XSPEC}) as  soft photon spectra. 
In fact, \citet{BRT99}, hereafter BRT99, demonstrated that in a limited energy 
range (for example {\it RXTE}/PCA) the {\sc diskbb} is fit by 
a {\sc bb} spectrum. The color BB temperature is the best-fit parameter 
which is related to the effective radius of the emission area of the disk. It is 
evident that  the best-fit  values of the color temperature and effective radius 
depend on the energy range of the instrument (see BRT99).
Any disk model requires the integral of the local spectra over radius. 
These local emergent spectra are not necessarily blackbody-like spectra and
they are indeed formed as a  result of the radiative transfer effects 
which depend on the assumption of the local density and temperature structure.
Moreover the specific diskBB model assumes a temperature distribution with 
the disk radius that is likely not the real distribution.
Thus, taking all that into account, we have preferred to describe the
thermal ({\sc bb}--like) component of the source spectrum by adding 
another {\sc bb} to that {\it already} included in the BMC model.

Following XSPEC notation, the model used to describe our spectra is given by
$F(E)= WABS\times(BBODY+ BMC)\times SMEDGE \times HIGHECUT$. Here 
{\sc wabs} is the XSPEC photoelectic 
absorption model which takes into account a possible intrinsic photoelectic 
absorption in the source and/or external absorption of X-ray source emission   
in the way towards the Earth.
The $BMC$ Comptonization model describes both direct (from an innermost part 
of the disk) and Comptonized photons (from the corona), the added {\sc bbody} 
component describes the direct soft photons belonging to  the disk part  at a 
greater distance  from the central object which spectrum can be only slightly  affected 
by upscattering in the corona. The {\sc smedge} (smeared K-edge) model  represents 
a possible interaction of  the photons originated in the inner  part of the 
X-ray source with a wind \citep[see][]{lt07}.  The {\sc highecut} (high-energy cutoff) 
model enables to take 
into account the curvature effect of the Comptonized photon spectrum  at 
high energies (that is above 100 keV in the LHS spectra) due to the recoil effect 
which is neglected  by BMC. {\sc highecut} is found to be needed to 
describe the LHS energy spectrum, while, for the HSS, no high energy 
cutoff is needed.

Most of the 39 spectra are found to be very well fit by the assumed input model. Only
for 9 spectra the model  has a null hypothesis probability less 
than 0.05, with 3 of them with  null hypothesis probability less than 0.01,
the smallest value being $3.7\times 10^{-3}$.

The best-fit parameters of 6 typical spectra extracted from the
entire set are shown in Table~~\ref{tbl-2}. 
As can be seen, the temperature $kT_{sbb}$ of the soft blackbody 
varies from $0.18$ to $0.36\,$keV, while that of the BMC, $kT_{bb}$, varies
in the range from $0.39$ to $0.56\,$keV, depending on the spectral state.
For the same time intervals of Table~\ref{tbl-2}, we show in Fig.~\ref{sp_evol}
the best fit $EF(E)$ spectra in keV~cm$^{-2}$~s$^{-1}$ units. 
In the LHS spectra I105 and I111
the relevant Comptonization contribution is related to the thermal 
Comptonization of soft (disk) photons  off hot electrons of Compton 
cloud (corona). Instead, in  the HSS spectra (see 
spectra I215, I216, I303, and I306 in Fig.~\ref{sp_evol}),  the blackbody emission
(both {\sc BBs}), presumably related to the intrinsic disk spectrum, is dominant 
in the resulting 
spectrum, while the hard (steep {\sc pl}) photons, probably  formed in the 
relatively 
cold (a few keV temperature) compact region, are only a small fraction of 
the total flux. 

As we noted above, the use of the thermal Comptonization model (COMPTT) 
is physically justified for the fitting of the LHS spectra  only. 
For comparison with the results reported in Table~\ref{tbl-2}, for time
intervals I105 and I111 we implemented a fit using a model $WABS\times(BB+COMPTT)\times SMEDGE$ (see Table~\ref{tbl-3}).
In  fact, the fit quality  obtained using the COMPTT model are not so 
good as that in the case of BMC [compare the $\chi^2/({\rm dof})$ values 
in Table~\ref{tbl-2} 
and \ref{tbl-3}] {presumably because a modest converging flow contribution
to the overall Compton upscattering is already present in TOO1}.
In general, the worse  
quality fits introduce  additional systematic errors in the 
best-fit model parameters.
In particular, the photon index,  obtained using COMPTT best-fit optical depth $\tau$ and electron temperature $kT_e$ and using the formula for the index related to $\tau$ and $ kT_e$ [see \cite{tl95}],  
is 0.4 higher than that inferred using BMC.

In the sections  4--5 we present the arguments for the compactness of the 
steep {\sc pl} emission region which is presumably {\em a converging flow} 
streaming towards the {\em innermost part of the BH}.  

\bigskip
\subsection{Observational evidence of the converging flow (black hole) in XTE J1650-500}
\smallskip
It is important to emphasize once again (see above) that the spectral index $\alpha$ (=$\Gamma-1$) is 
the reciprocal of the Comptonization parameter $Y$ [see e.g. \cite{btk07}] and it is  independent of the  
Comptonization model (thermal or bulk). The parameter $Y=N_{sc}\eta$ is 
the mean number of scatterings $N_{sc}$ times the average fractional energy change per scattering
$\eta$.  The number of scatterings of Comptonized (upscattered) photons  in the converging flow (CF)
$N_{sc}$  is proportional
to the optical depth $\tau$, given that in the converging flow the photons can be only effectively 
upscattered in the direction of the bulk motion, and thus   $N_{sc}\propto R/l=\tau$, where $R$ is 
the CF characteristic radius and $l$ is the photon free path. It is also worth noting that, in the 
converging flow, $\tau$ is also equal to the dimensionless mass accretion rate $\dot m=\dot M /\dot M_{\rm Ed}$  
where  $\dot M_{\rm Ed}$ is defined as $\dot M_{\rm Ed}=L_{\rm Ed}/c^2$  [see \cite{bmc} for more details]. 
On the other  hand,  the change of the CF fractional energy per scattering $\eta$ is inversely 
proportional $\tau$ $(\eta\propto1/\tau)$ when $\tau\gg1$ [see for more details in \cite{lt07}].

Consequently $Y=N_{sc}\eta$ and thus $\Gamma=1/Y+1$ saturate to a constant value when $\tau$ (or  
$\dot m $) increases.   Observationally this index saturation can be revealed if one can find a
correlation between $\Gamma$ and BMC normalization $C_{N}=L_{39}/d^2_{10}$ where
$L_{39}=L_{d}/ (10^{39}~{\rm erg}s^{-1})$ is the luminosity of the seed photon source (disk) 
in units of $10^{39}$ erg s$^{-1}$ and $d_{10}$ is the distance to source in units of 10 kpc.  
This correlation  of the index $\Gamma$ with  $C_N=L_{39}/d^2_{10}$ implies the correlation of $\Gamma$
vs $\dot m$ given that $L_{d}$  as a disk luminosity  is proportional to $\dot m$, 
namely $L_d\propto \varepsilon_{eff} m\dot m$, where $m=M/M_{\odot}$ is the black hole mass in units 
of solar mass and  $\varepsilon_{eff}$ is the efficiency of the  gravitational energy release  
in the accretion disk [see e.g. \cite{ss73}]. In principle,  the efficiency  $\varepsilon_{eff}$ can 
depend on the inner radius of the disk $R_{in,d}$, i.e. indirectly depends on $\dot m$. However,
in HSS, $R_{in,d}$ presumably reaches its lowest limit of about $3R_{\rm S}$ [see e.g. \cite{st99}] and 
thus, in HSS, the index vs. $C_{N}$ correlation is equivalent to the correlation 
between index and $\dot m$. This  implies that in HSS the index saturation with $C_{N}$ is 
equivalent  to the index saturation with $\dot m$.

In Figure \ref{index_vs_bmcnorm} we present an observational evidence of the index 
saturation with  $L_{39}/d^2_{10}$, and thus with $\dot m$,  which can be considered as 
{\it an observational  signature of a converging flow (black hole) in XTE J1650-500}.


\subsection{On iron fluorescence  line determination}

Limiting our considerations to the \sax\ TOOs  we confirm the
broad excess at $\sim6.4$ keV found by M04 in all of the three 
observations if  we adopt, as M04 did, the XSPEC reflection  model {\sc pexriv}. 
However, the excess is present only 
in the first TOO if we use  a two--phase accretion disk corona model which 
also includes the reflection effect (XSPEC {\sc compps} Model) (Montanari et al. 2008, in 
preparation). 
It is worth noting that when we use COMPPS model   we find a strange effect 
(unexpected in the scenario described by this model) that the iron line 
(thought to be due to reflection from the disk) is present when the reflection 
parameter is low (in TOO1) and it is completely absent  when the reflection parameter 
becomes larger (in TOO2 and TOO3) \citep[see][]{mon}. Also as we have already 
mentioned above,  in the framework of the COMPPS model  there is 
a puzzling effect of significant decrease of    the optical depth of the 
Compton cloud when the source evolves to the HSS \citep{mon}.

On the other hand, using the input continuum model described in 
Section~\ref{s:spectra} 
and the entire \sax\ spectral energy range {\it we find a very marginal 
evidence for the presence of the Fe fluorescence line in the data}.
We illustrate  our findings in Figure \ref{range}.

Adopting the 0.12--200 keV energy range,  
the fit of the continuum model  $WABS \times (BB+BMC) \times SMEDGE  
\times HIGHENCUT$  to the spectrum I108, used as an example,  is shown in the 
panel 2 (starting from the top).  
As can be seen, the above continuum model gives a very satisfactory description 
of the data ($\chi^2/{\rm dof} = 150/164)$, with no need of adding 
a Fe K$_{\alpha}$  line model. Instead the $SMEDGE$ model appears crucial 
for a good fit. The fit of the above continuum model with 
no $SMEDGE$ to the same spectrum (see top panel)
is highly unacceptable ($\chi^2/{\rm dof}=430/165$). The apparent
sinusoidal pattern of the residuals to the model is 
a clear sign of the bad fit quality.

Note the addition of a K$_{\alpha}$ line ({\sc laor} model) to the
continuum model with no $SMEDGE$ gives an unacceptable fit
($\chi^2/{\rm dof} = 300/163)$ (see panel 3), but it becomes acceptable 
($\chi^2/{\rm dof} = 136/109$, see panel 5), if we limit
the spectrum energy bandwidth  to 1.5--60 keV (the band used by M04)
\footnote{Note that a fit of the model  $WABS \times (BB+BMC) \times HIGHENCUT$
with no Laor component does not fit the data in this limited energy band
(see panel 4).} . 
However the model which  is valid in 1.5--60 keV, is not valid in
the entire \sax\ energy band. Namely the model 
$WABS \times (BB+BMC+LAOR) \times HIGHENCUT$, with the best-fit parameter values 
found for the 1.5--60 keV energy band, does not fit the data
related to  the entire \sax\ energy range
from 0.12 to 200 keV ($\chi^2/{\rm dof} = 1315/164$, see panel 6).

We have also attempted to understand why M04, using the $SMEDGE$ model as we do,
obtained different results. Limiting the analysis to the  spectrum I108, 
assumed as an example, we actually find that, freezing the energy of the Edge to 
the best-fit value reported by M04 [$E_{edge}=8.3$~keV {being consistent with a} 
very high ionization state of iron ($ion~ XXII$), see \cite{k04}] gives the 
presence of a significant emission line. The line disappears when $E_{edge}$ 
is left free to vary, finding, as a best-fit  value, 7.2 keV.

To conclude, iron  K$_{\alpha}$ line  appearance strongly depends on the
energy band used to describe the continuum emission.  With our best-fit continuum
model, that gives the best description of the source spectrum in the 
entire 0.1--200 keV \sax\ energy band,  {\it the line is not required}. 

Although no significant emission line is evident in the data, adding the  
XSPEC {\sc laor} model to our best-fit model (see Section 3) slightly improves the fits 
($\Delta\chi<10$ with a decrease of 4 dof for about 160 initial dof) for some observation
intervals belonging to TOO1 (LHS).
The resulting line, of equivalent width ranging from few tens eV to about 100 eV, 
{\em is almost symmetric} with its center being 
situated in the energy range between 5 and 5.5 keV. 
For all of other intervals the normalization of the line goes to zero.

Concerning the SMEDGE model used in the best fit description of our spectra,
it is worth noting there is a relation between Thompson optical depth $\tau_{\rm T}$ 
of the iron emission area 
and the SMEDGE  maximum absorption optical depth $\tau_{max}$ at threshold 
[see \cite{basko} and ST06]:
\begin{equation}
\tau_{max} = (Y+Y_0)\,\tau_{\rm T}\,\left(\frac{7.8\,{\rm keV}}{E_{th}}\right)^3
\label{tau_max-tau_T}
\end{equation}
where $Y$ and $Y_0$ (in units of the cosmic values) are the abundances of 
the elements with a charge $Z<26$ and the iron abundance respectively, while 
$E_{th}$ is the K--shell ionization threshold energy. In our case 
$Y\simeq 1$, $Y_0\simeq 1$, $E_{th}=7.1$ keV, so
\[
\tau_{max} \simeq 2.65 \tau_{\rm T}
\]
For instance, $\tau_{max}= 2, ~3,~ 4,~ 5$ correspond to   $\tau_{\rm T}=0.75, ~1.1, ~1.5, ~1.9$ 
respectively.

We find that, in the LHS, the Thomson optical depth $\tau_{\rm T}$    is relatively low, i.e. 
about 0.7 whereas, in the HSS,  $\tau_{\rm T}$ is about 2 (see 
Table~\ref{tbl-2} for the values of $\tau_{max}$  and Eq. \ref{tau_max-tau_T} for the
$\tau_{max}-\tau_{\rm T}$ relation).
Thus, while in the LHS the possible detection of the line is consistent with the
value of the Thompson optical depth $\tau_{\rm T}\sim 0.7$ (see Table \ref{tbl-2}), 
in the HSS, its non detection is consistent with the smearing due to a significant
Thompson optical depth $\tau_{\rm T}$ in the range between $\sim 1.3$ and 
$\sim 2$ (see above).

One can ask a  fair question on the relation between  the K-edge and K$_{\alpha}$ 
appearances  in the data. In many  models  K$_{\alpha}$-line formation  is 
indeed  strongly related to edge formation [see e.g. \cite{bst74}, 
\cite{basko}, \cite{k04}, \cite{lt04}, \cite{lt07}].  However it is 
not always true that the detection  of the  K-edge in the data implies  
the K$_{\alpha}-$line detection.  
The physics of the K-edge and K$_{\alpha}$ formation is the following: photons 
of energies higher than 7.1 keV ionize the iron K-shell and this ionization 
effect  leads to emission of K$_{\alpha}-$photon (iron K$_{\alpha}-$fluorescence) 
with a probability of $\omega_K\lax0.3$ (where $\omega_K$ is K-yield),  whereas 
with probability  $(1-\omega_K)$ the K$_{\alpha}$  energy goes to on ionized 
electron.  So the edge can be always there but the K$_{\alpha}-$photon is 
emitted with a probability less than 0.3 and in addition this 
probability is affected by the electron density of the surrounding 
medium (see more details in the aforementioned references on this 
subject). 
Even if the K$_\alpha-$photon is emitted it does not mean that this photon 
can be detected and seen by the Earth observer. These photons can be 
absorbed when they propagate through the medium (for example in the wind) 
and  they can be also scattered off electrons and  significantly change 
their energy, i.e. washed out from the resulting spectrum. All these 
effects of line appearance and disapperance have been reproduced in 
simulations by  \cite{lt04} and \cite{lt07}.

\section{Power spectrum and timing-spectral correlation}

We succeeded to derive
accurate power spectra (PS) of the source time variability thanks to the high 
source brightness  during outburst. In particular, 
during TOO-1, when the source was mainly in the LHS, it was possible to obtain the PS 
in several energy intervals within
the broad 0.1--200 keV \sax\ operational band. 

\subsection{Phenomenological (Lorentzian) Model}

First  we made {the} analysis of  {the power spectra} using,  as an input model, 
the superposition of Lorentzian functions (Lorentzian model). 
We compared their characteristics  with { those} identified by
KVK08 for a number of  BHs.
A Lorentzian function is given by  
\begin{equation}
L(\nu)= K \frac{\hat\Gamma/2\pi}{(\nu -\nu_0)^2 + (\hat\Gamma/2)^2}
\label{lorentzian}
\end{equation} 
where $\nu_0$  is the centroid frequency and $\hat\Gamma$  is the  full width at half maximum
of the function (FWHM).
 
{We used broad Lorentzians to fit  the continuum spectrum  and narrow Lorentzians 
to fit  QPO features.} 
In Figure \ref{hard_vs_soft_PDS} ({\it bottom panel})  we illustrated the fitting 
results and the characteristics of the PS components  
in terms of frequency$\times$power diagram. Note the  power spectrum unit 
is $(rms/{\rm mean)^2/Hz}$ and thus  
the frequency  $\times$power unit is $(rms/{\rm mean})^2$.   
Below we presented  a comparison of these 
characteristics  with those  found by  KVK08.   
  
{ The continuum of all derived LHS PSs is found to be well described  by the
sum of two broad Lorentzians { $L_1$ and $L_2$. Following the KVK08 notation, 
we formally  call them $L_b$ and $L_h$, respectively. We report the best fit 
characteristic frequency  of these broad Lorentzians in Table~\ref{tbl-lorentzian}. 
 This frequency corresponds to the maximum of
the $\nu L(\nu)$ diagram $\nu_{max}= [\nu_0^2 + (\hat\Gamma/2)^2]^{0.5}$.
The presented PSs are  results of integration over  the entire duration of each TOO.

Using  Fig. \ref{hard_vs_soft_PDS} and Table~\ref{tbl-lorentzian} one can see  that, 
{in addition to the two broad Lorentzians $L_b$ and $L_h$,
the 2-10 keV  power-frequency diagram  of  the LHS (TOO1) requires  
two narrow Lorentzians: a low frequency (LF) QPO and a higher frequency QPO 
at  $\nu_{QPO1}\sim 1.5$~Hz and   $\nu_{QPO2}\sim 3$ Hz, respectively.} 
These two QPOs were introduced   by KVK08  as  LF and  LF$^+$, respectively.
Note  that $\nu_{QPO2}\sim 3$ Hz  (or  LF$^+$) was  found to be consistent 
with the second harmonics of   $\nu_{QPO1}$ (or LF). 

However, in the 0.1--2 keV energy channel, no evidence of $\nu_{QPO2}$ is found 
and the  characteristic frequency of the broad Lorentzian $L_h$, that describes
the hump to the right of the QPO frequency (see blue line in the
right bottom panel of Fig.~\ref{hard_vs_soft_PDS}), is at  23 Hz. Note the $L_h-$hump is located at 
 $\sim 2$~Hz in the 2--10 keV power-frequency diagram}. 

%
%
In Table~\ref{tbl-energy_depend_Lorentzian}  we show the energy dependence  of the 
LHS PS components  in the broad energy bands from 0.1-2 keV to 15--200 keV.
The characteristic frequencies of $L_b$ are consistent with each other 
up to 15 keV and the $L_h$ characteristic frequencies are consistent
with each other in the 2--4, 4--10 and 8--15 keV energy intervals. 
Using the $L_b-L_h$ relation found 
(for the energy band 2-25 keV) by Wijnands and van der Klis [\cite{wk99}],
 updated in KVK08, we find that our medium energy 
data points are consistent with that relation. 
Also the PS {narrow} components L$_{QPO1}$ and  L$_{QPO2}$ clearly detected in the 4-10 keV and 
8-15 keV  bands  satisfy the Wijnands and van der Klis $L_{LF}-L_b$ relation (see
Fig. 10 in KVK08).

It is worth noting that,  in  the 0.1--2 keV and  15-200 keV energy intervals,  the $L_h$  frequencies  
are an order of magnitude higher than those at medium energies. In fact, it is the first time that 
the high values of $\nu_{max}$ ($\sim23$ Hz, $>50$ Hz  in the  0.1--2.0 keV and 15--200 keV 
energy bands respectively),  are discovered from this source. Their identification with the 
$L_h$ introduced by, e.g., by KVK08, is not obvious. 
Unfortunately, there are no data in the literature to statistically study a
possible correlation between $L_b$ and $L_h$ at these energies, unlike what can be done  
at medium energies with RXTE data (see, e.g., KVK08). The 
$L_h$ components in   the 0.1--2 keV and 15--200 keV 
energy bands can be related to PS physical components which are different from 
that in the medium energy bands (see  details of the physical interpretation in  section 4.2).

{ In  the HSS, the PS continuum is still found to be 
well described  by the sum of two broad Lorentzians, but in this case 
(see Table \ref{tbl-lorentzian} and Fig.~\ref{hard_vs_soft_PDS}), $L_h$ is no more 
visible, while a very low frequency noise (VLFN) emerges.} 
It is important to emphasize that the characteristic frequency of L$_b$  increases 
from the LHS to the HSS. 
Our finding is in agreement {with the fact} that the L$_b$ characteristic 
frequency is shifted to higher values 
during the source evolution from the LHS to the HSS [see  also  Fig. 4 in KVK08].
We also confirm the KVK08 result that LFVN component occurs in the HSS, but  
is absent in LHS [compare our Fig. \ref{hard_vs_soft_PDS} ({\it bottom panels}) 
with panels {\it a} and {\it e} in Fig. 4 of KVK08].

Furthermore we find that the Q-value, { defined} as
$Q=\nu_0/ \hat\Gamma$, 
is in the range of 0--0.5 for L$_b$, L$_h$ components and 
that is in the range of 4--6 for PS narrow components  
$L_{QPO}$. 
As one can see from Fig. \ref{hard_vs_soft_PDS} that  $rms/{\rm mean}$ values 
are  10\% and  30\%  for  L$_b$ and L$_h$ of the LHS power spectra  
components respectively in 0.1-2 keV. The corresponding {$rms/{\rm mean}$  values  
for the  LHS L$_b$ and L$_h$ in 2-10 keV band  are about 10\% 
(see left bottom panel of Fig. \ref{hard_vs_soft_PDS}).   However the 
variability power of the PS components   drastically decreases in the HSS, 
and their rms-values  are less than 1\%. 
It is worth noting that these ranges of Q and rms values are consistent with that 
found by KVK08 (see Table 1 there).

\subsection{Physical (diffusion) model} 

The analysis of the PSs was performed using  a simplified version of the  
diffusion model {(see TSA07 and TS08)} in which the PS continuum shape 
at frequencies below the driving frequency can be approximated by 
a broken power-law ({\sc bknpl}). The results of the model fitting to 
the data are shown in Tables \ref{tbl-4}-\ref{tbl-6}.  
We also present the model fitting results in 
Figs.~\ref{hard_vs_soft_PDS}-\ref{pds_vs_energy}.

In particular, in Fig.~\ref{pds_vs_energy} and Table~\ref{tbl-4} we show 
the  energy dependence of the LHS PSs and  the 
variability power, that decreases with energy (mainly above 10 keV), while it 
preserves the self-similar shape for energies higher than 2 keV.
These results obtained for  the LHS  have relevant implications.
The self-similar shape of the $>2$ keV spectra implies that photons of 
these energies are  produced (upscattered) in the same geometrical configuration.  
Following TSA07, the break frequency found in the LHS PSs
is related to a diffusion time of  perturbation propagation 
while the QPO low frequency is an eigenfrequency of the volume 
(magnetoacoustic)  oscillation of the medium (in our case it is a 
Compton cloud). Our results show that
the values of these PS characteristics are similar  in all PSs  for photon  
energies  higher than 2 keV. 

As we have already demonstrated in Section 3,  the energy spectral behaviour 
(Fig.~\ref{sp_evol}), even within
TOO-1, shows a slight softening  from the time interval I105 to I111. 
In order to see
the effect of this softening in the PS, we have compared the 4--10 keV PS derived 
in the time interval from I101 to I107 with that related  to the time interval 
from I109 to I112 in the 4--10 keV energy band. Note that the 4--10 keV band was chosen
because  we find  the best signal to noise 
ratio {\it in this interval}.

The result is presented in Table~\ref{tbl-5} and in Fig.~\ref{pds_shift}.  
The main  effect   shown in  Fig.  \ref{pds_shift} is  the shift to
higher frequencies of the PS which correlates with  the photon spectral softening 
{(see Fig.~\ref{sp_evol})}. 
Independently of the continuum adopted model 
the LF QPO centroid frequency 
increases while the continuum noise 
remains stable in its shape. For the diffusion model continuum, as can be seen from 
Table~\ref{tbl-5},  the LF QPO  frequency 
increases from $1.47\pm0.03$~Hz to 
$1.66\pm0.08$~Hz (uncertainties at 90\% confidence level).


In order to illustrate the evolution of the power spectra from TOO-1 
to TOO-2, when the energy spectra undergo their  transition from the  LHS to
the HSS (see I215 and I216 spectra  in Fig.~\ref{sp_evol}), we 
compared the LHS PS with the HSS PS  in two energy bands where the signal is
highest: 2-10 keV and 0.1-2 keV  (see { top} left and { top} right panels of 
Fig.~\ref{hard_vs_soft_PDS}, respectively).
As can be seen from this Figure, the relative power of  the HSS PS (with respect 
to the LHS PS) and its shape strongly depend on the energy band.  

We find { (see Table~\ref{tbl-6})}  that the 0.1--2 keV HSS PS is well described by
the sum of a {\sc pl} plus a {\sc bknpl}, with no evidence of QPOs
in contrast to  the 0.1--2 keV LHS PS  which is  described 
 { (see Table~\ref{tbl-4})  by a  {\sc bknpl} plus 
a high ($\nu_0 \sim 14$~Hz) frequency broad  Lorentzian (FWHM$\sim 40$~Hz).
On the other hand   the 2--10 keV HSS PS 
(see Table~\ref{tbl-6} and Fig.~\ref{hard_vs_soft_PDS})
can be fit with  a {\sc pl} with plus a constant, { while the 2--10 keV LHS PS
is fit with a {\sc bknpl} plus two narrow Lorentzians, one of which is consistent
to be the second harmonics of the other at $\nu_0 \sim 1.5$~Hz. 

%
These results confirm what  was previously found by \cite{h03}
using {\it RXTE} data, that there are  prominent QPO features in 
XTE J1650-500  PSs [see also \cite{rs01} and \cite{w01}].

In  the LHS, the Comptonization component is dominant and thus we should  
detect the X-ray emission along  with  its variability related to the 
Comptonizing region (Compton cloud) only. Indeed in all of the LHS PSs 
shown in  Fig.~\ref{hard_vs_soft_PDS}, the {VLFN} component is not seen.  
{\it Its absence 
is in agreement with the low intensity level of the direct  disk blackbody component 
in the  LHS energy spectra} (see  I105 and I111 spectra in Fig. \ref{sp_evol}).
It is important to emphasize  that, in the LHS,  the typical values of the 
characteristic frequencies  of the {VLFN} and {\sc bknpl} 
components are of the order of $10^{-7}-10^{-6}$ Hz and $0.1-1$ Hz, respectively 
(see TSA07 for Cyg X-1 case and KVK08).} Thus, in this spectral state,
the disk variability  can be clearly detected  only  if  frequencies  lower than $10^{-3}$ Hz 
corresponding to photon energies less than  $\sim 0.1$  keV (see TSA07).
However this is  not the case for \sax.

On the other hand in the HSS PSs the VLFN component appears 
(with the  power depending on the energy range) along with the {\sc bknpl} component. 
Note that this latter component is barely visible  in the 2--10 keV PS 
(see Fig.~\ref{hard_vs_soft_PDS}). 
This result is in  agreement with a low contribution  of the high energy tail in 
the HSS energy spectra (see Fig. \ref{sp_evol}).
Following TSA07, we can interpret the 
VLFN component observed in the HSS as related   to the timing  response of  
the extended disk, while {\sc bknpl}  or $L_b$ (in terms of Lorentzian model)  is  associated with mass 
accretion rate perturbations in the Compton cloud (see the HSS PS  in the 
right panels of Fig.~\ref{hard_vs_soft_PDS}).


We find  that  in the 0.1--2 keV band  not only   $\nu_{b,max}$ (using the Lorentzian model) but 
also $\nu_{br}$ (using the diffusion model)   increase from TOO--1 (LHS) to TOO--2 (HSS)  
by a factor 60 ($\nu_{b\ max}$ increases from $0.59_{-0.13}^{+0.06}$ Hz to $36\pm 3$ Hz, 
while $\nu_{br}$ increases from $0.28\pm 0.03$ Hz to $\sim 17^{+7}_{-2}$ Hz).
%
%
This high shift of the related  frequencies  is in agreement with the corona (Compton cloud) contraction during the 
LHS-HSS transition (see TSA07).
Searching in the literature,   we found that in Cyg X-1 PSs   the break frequency  changes from 
$\sim 0.03$ Hz to    $\sim 18$ Hz value  when photon index varies from 1.5 to 2.1  (see Fig. 8 in ST06).
Also we should point out  that \cite{wk99} (see Fig. 2) reported the range 
for the break frequency in BHs is  $0.03-30$ Hz. 

The $\sim 14$ Hz hump clearly seen in the 0.1-2 keV LHS PS (see 
Fig.~\ref{hard_vs_soft_PDS} and Tables~\ref{tbl-lorentzian}, \ref{tbl-4}) 
disappears at higher energies 
presumably because of the lack of statistics (compare with Fig. \ref{pds_vs_energy}). 
TSA07 and TS08  give an explanation of this hump in the  
high frequency tail of the power spectrum  of Cygnus X--1, in terms of 
driving frequency  of the perturbation oscillations of the Compton cloud.

\section{Discussion and Conclusions}

In this Paper we present the results of simultaneous  spectral  and timing 
analysis of the BHC XTE J1650-500 observed by \sax.

The energy spectra are well described by the sum of a Comptonization and 
soft blackbody components along with a smeared edge.  
In particular,   we  argue that the iron line  appearance strongly depends on the
energy band used to describe the continuum emission. Using both the high energy
($>$20 keV) and the low energy ($<$1.5 keV) bands, which are both
crucial to describe the real continuum level, the presence of the iron line feature is not
mandatorily required. 
Thus we conclude  that the line feature is not so prominent in  \sax\ data  as 
initially reported by M04. 
One may also  attribute the difference of  our conclusion regarding the presence  of 
the K$_{\alpha}-$iron line in the \sax\ data and  that by M04  to  a different 
treatment of systematic errors and different  procedure of  data integration over time. 

We  find the clear evidence of an evolution of the spectra from the LHS to the HSS 
(see Fig. \ref{sp_evol}).
 In the LHS the Comptonization component dominates, in contrast to  the HSS when 
the blackbody component dominates the emergent spectrum. 
Moreover in the  \sax\ data (see  Fig. \ref{index_vs_bmcnorm}) we find a strong signature 
of the index saturation with BMC normalization $L_{39}/d^2_{10}$, which is, in fact,
proportional to the disk mass accretion rate}.   {\it This index saturation vs. mass accretion rate
can be considered as an observational evidence of the converging flow (black hole)} in XTE J1650-500 
[\cite{TZ98} and \cite{TF04} hereafer TF04].

The relative contribution of the 
Comptonization component (in the resulting spectrum) decreases when 
the source undergoes the hard-to-soft spectral transition 
(see Fig.~\ref{sp_evol}).  We interpret this as a result of  
contraction of the Comptonization region (Compton cloud) during this
spectral evolution [see \cite{TLM98}, hereafter TLM98, and TF04].

We can infer this contraction effect  using our best-fit model 
parameters. One of the parameters of the BMC model 
is the normalization of the blackbody (BB) component
$C_N$ which is directly related to the effective area $A^{eff}_{bb}$ of 
the BB emission [$A^{eff}_{bb} \propto C_N/(T_{bb}^4\,(1+A))$]. 
If we multiply this effective area with the fraction of Comptonized 
component $f = A/(1+A)$, we obtain the effective area of the high energy X--ray 
emission region (the Compton cloud). Using the best-fit values for the LHS and HSS 
spectra we find that the area of the Compton cloud  decreases by a factor
five during the transition  from the LHS to the HSS (see Fig. \ref{sp_evol_area}).

Furthermore in Fig. \ref{comp_fractevol_vs_index} we demonstrate how the Comptonization
fraction $f$ decreases  when the photon index increases.
This correlation  presents one more argument for a Compton cloud contraction 
during LHS-HSS evolution.
While in  \S 4 we  argued  this contraction effect  using the power 
spectrum analysis only, here we   find more arguments for this contraction
using the energy spectrum analysis. 

Actually the power spectra follow the energy spectral evolution. 
The PSs related to  the softer energy spectra are weaker  and shifted to 
higher frequencies with respect to those related to the harder spectra.  It 
leads us to the conclusion that the emission area (CC) does contract 
when the source undergoes the hard-to-soft state  transition. 

Furthermore TLM98 argued that the size of CC  strongly depends 
on the Reynolds (${\rm Re}$) number of the accretion flow, and not on just the mass 
accretion rate.  
In other words, the LHS-HSS transition should be  dictated by Re-number values, 
which is low in the LHS 
and high in the HSS, and  the CC size decreases when Re-number increases.  TS08 
check this TLM98 prediction  using Cyg X-1 observations by {\it RXTE}. First,  
they find a  method to infer the Re-number of the accretion flow using 
the observations. Then they demonstrate that when Re-number increases 
all characteristic frequencies of the power spectrum are indeed  shifted  
to higher values as a result of  Compton cloud contraction during  
the LHS-HSS transition.

It is well known  from the observations  of X-ray BH binaries 
[see e.g. \cite{wk99}, ST06, ST07 and  KVK08]  that  the power spectra evolve along  with 
the photon spectra.
Here we find that our results for XTE J1650-500 are related to the findings 
by ST06 for Cyg X-1, 
where the authors, analyzing  $\sim10$ years of observational  data 
from {\it RXTE} archive, found strongly  correlated  characteristics 
of  power spectra  and photon spectra (photon index).
We have to admit that our claims of temporal and spectral correlation in  
XTE J1650-500 are based on  only the limited  set  of the BeppoSAX    data,  while  the 
results of the extensive  {\it RXTE} data set  for XTE J1650-500  by  
\cite{st08}, hereafter ST08, do 
indicate a strongly correlated temporal and spectral evolution in this source. 

This  suggests that the underlying physics of the radiative and 
oscillatory processes  is common  in  X-ray BH binaries.
The consistency of the Comptonization model
also suggests that their  spectra emerge  from similar Compton cloud  
geometric  configurations whose  sizes shrink during the state transition.
The disk becomes more powerful at softer states and the cooling of 
Compton cloud is mostly dictated by the strong soft disk emission. The 
photon field and surrounding plasma are almost in equilibrium: the 
plasma temperature is the same order of the photon blackbody color 
temperature (of a few keV).
The converging flow located in the innermost part of compact cloud 
is the only place where photons can be upscattered to energies 
of order of $m_ec^2$ and higher due to the dynamical Comptonization.

In the HSS PSs we  see the presence  of two noise  
components: VLNF and {\sc bknpl}  (or $L_{VLFN}$ and $L_b$ in terms of Lorentzian model). We suggest that the VLFN
component is associated with the extended  disk and {\sc bknpl}   is  
presumably  formed as a result of the diffusive propagation of the mass 
accretion rate perturbation  in the compact  Compton cloud.  In the LHS PSs 
only {\sc bknpl} 
is observed, while the VLFN related to the
disk component, presumably very weak
and at very low frequencies, cannot be observed in the \sax\ data.
 
We also show that in XTE 1650-500  the power of X-ray variability drastically 
decreases towards   the HSS. Note this effect   have been  known for a number of BHs 
(see KVK08 and TS08 for details of the observations and explanation of 
this effect, respectively).
The relative power  of the HSS PS with respect of that in 
the LHS clearly depends on  the energy bands.  We present (for the first 
time in the literature) energy-dependent PSs of X-ray emission using 
\sax\ data. While we find a noticeable dependence of the
PS strength on energy,  the PS shape is found to be almost self-similar 
in different energy bands (at least for  the LHS PSs). 

Moreover,  we show that the 
break and QPO low frequencies of  the XTE J1650-500  PSs  which we derived,    
correlate with the spectral state, in particular with the {\sc pl} 
photon index of the energy spectrum. We succeed  to determine the LF QPO  
corresponding to two different LHS spectra (indices):
$\nu_{QPO}=1.47 \pm 0.05$ Hz,    $ 1.66 \pm 0.08$ Hz related 
to  $\Gamma=1.74\pm 0.01$  and      $1.79 \pm 0.06$ respectively.
Thus our data points based on \sax\ data for XTE J1650-500 are consistent  
with the index-QPO correlation found by  ST08, based 
on the  {\it RXTE } data for   XTE J1650-500.

\acknowledgments
We would like to thank  anonymous referees for constructive suggestions 
to  significantly  improve the paper  presentation.

\clearpage
%
%
\begin{table}
\begin{center}
\caption{Log of the three observations of XTE J1650-500\label{tbl-1}}
\begin{tabular}{ccccccc}
\tableline\tableline
Obs. & Start time (UT) & End time (UT) & LECS & MECS & HPGSPC & PDS \\
 & & & ks & ks & ks & ks\\
\tableline
1 & 2001 Sep. 11 10:57:04 & 2001 Sep. 12 18:43:06 & 20.3 & 47.4 & 47.0 & 22.2\\
2 & 2001 Sep. 21 18:15:39 & 2001 Sep. 23 14:03:40 & 15.9 & 63.9 & 73.3 & 30.3\\
3 & 2001 Oct. 03 16:36:53 & 2001 Oct. 04 10:15:56 & 7.2 & 27.8 & 33.0 & 12.8\\
\tableline
\end{tabular}
\end{center}
\end{table}

\clearpage
%
%
\begin{deluxetable}{llllllll} 
\tablecolumns{8} 
\tablewidth{0pc}
\rotate 
\tabletypesize{\footnotesize}
\tablecaption{Parameters of the model: $WABS\times(BBODY+BMC)\times SMEDGE\times
HIGHECUT$. $HIGHECUT$ was used only for the LHS. Parameters not listed in
the table were held fixed. For $HIGHECUT$: $E_{cut}=10$ 
keV; for $SMEDGE$: $E_{edge}=7.1$ keV, index for photo-electric absorption = 2.67
(default value), smearing width = 10 keV. Errors are 
given at the 90\% confidence level.}
\tablehead{
\colhead{Model} & \colhead{Parameters} & \colhead{I105} & \colhead{I111} & \colhead{I215} & \colhead{I216} 
& \colhead{I303} & \colhead{I308}
}
\startdata
{\sc wabs} &       &    &     &    &     &     &     \\                  
		& $N_{\rm H}$ ($10^{22}$~cm$^{-2}$) & $0.62^{+0.04}_{-0.07}$ & $0.62^{+0.08}_{-0.07}$ & 
$0.48^{+0.01}_{-0.02}$ & $0.49\pm 0.02$ & $0.50\pm 0.03$ & $0.52\pm 0.02$ \\
{\sc bbody} &       &    &     &    &     &     &     \\                  
	& $kT_{sbb}$ (keV) & $0.19^{+0.02}_{-0.01}$ & $0.19\pm 0.02$ & 
$0.36^{+0.03}_{-0.01}$ &  $0.35\pm 0.02$ & $0.34\pm 0.03$ & $0.33\pm 0.02$ \\
{\sc bmc} &       &    &     &    &     &     &     \\                  
	& $\Gamma$ & $1.74^{+0.01}_{-0.02}$ & $ 1.81\pm 0.02$ & 
	$2.50^{+0.02}_{-0.03}$ & $2.38\pm0.03$ & $2.32\pm0.08$ & $2.34\pm0.04$\\
 & $kT_{bb}$ (keV) & $0.40^{+0.02}_{-0.01}$ & $0.39\pm0.02$ & $0.56^{+0.06}_{-0.02}$
	 & $0.55^{+0.04}_{-0.03}$ & $0.52^{+0.03}_{-0.02}$ & $0.54\pm0.02$\\
 & $\log A$ & $0.30^{+0.08}_{-0.04}$ & $0.20^{+0.08}_{-0.07}$ &
	$-0.31^{+0.12}_{-0.05}$ & $-0.44^{+0.07}_{-0.06}$ &
$-1.06^{+0.09}_{-0.06}$ & $-0.83^{+0.07}_{-0.05}$\\
  & $N_{sbb}/N_{bb}$ & $0.8\pm 0.4$ & $0.9^{+0.9}_{-0.5}$ & $1.4^{+0.9}_{-0.4}$ 
	& $1.3^{+0.5}_{-0.3}$ & $1.1^{+0.8}_{-0.4}$ & $1.3^{+0.4}_{-0.3}$ \\
{\sc smedge} &       &    &     &    &     &     &     \\                  
	& $\tau_{\rm max}$ & $2.04^{+0.10}_{-0.13}$ & $1.96^{+0.17}_{-0.19}$ &
 $3.5^{+0.2}_{-0.4}$ &$3.6^{+0.3}_{-0.4}$ & $5.3^{+0.9}_{-1.0}$ & $4.5^{+0.6}_{-0.5}$ \\
{\sc highecut} &       &    &     &    &     &     &     \\                  
	& $E_F$ (keV) & $114^{+5}_{-6}$ & $114^{+8}_{-9}$ & --- & 
--- & --- & --- \\
          & $\chi^2$/dof &179.0/164 & 167.7/163 & 177.6/151 & 194.9/153 & 164.6/131 & 
	190.7/148\\ 
\enddata 
\label{tbl-2}
\end{deluxetable} 

\clearpage
%
%
\begin{table}
\begin{center}
\caption{Parameters of the model: $WABS\times(BBODY+COMPTT)\times SMEDGE$. Errors are given at the 90\% confidence level.
\label{tbl-3}}
\begin{tabular}{ccc}
\tableline\tableline
Interval & I105 & I111\\
$N_{\rm H}$ ($10^{22}$~cm$^{-2}$) & $0.38_ {-0.01}^ {+0.02}$ & $0.39_ {-0.02}^ {+0.03}$\\
 $kT_{sbb}$ (keV) & $0.30\pm 0.01$ & $0.31\pm0.01$\\
$T0$ (keV) & $0.45\pm0.03$ & $0.46^{+0.05}_{-0.04}$\\
$\tau$ & $1.33^{+0.06}_{-0.08}$ & $1.20^{+0.11}_{-0.16}$\\
$kT_e$ (keV) & $31^{+2}_{-1}$ & $32^{+4}_{-2}$\\
$\Gamma$ & $2.15\pm0.03$ & $2.12\pm0.06$ \\
$N_{sbb}/N_{comptt}$ & $0.52^{+0.02}_{- 0.01}$ & $0.68^{+0.02}_{-0.01}$\\ 
{\sc smedge}-$\tau_{\rm max}$ & $1.74^{+0.08}_{-0.12}$ & $1.7 \pm 0.2$\\
$\chi^2$/dof & 237.0/165 & 210.4/164\\
\end{tabular}
\end{center}
\end{table}

\clearpage
%
%
\begin{deluxetable}{llll} 
\tablecolumns{3} 
\tablewidth{0pc} 
\rotate
\tablecaption{Characteristics of BH XTE J1650-500 variability components of  the power spectrum 
for Low Hard State (TOO-1)  and High Soft state (TOO-2) [see  Fig. \ref{hard_vs_soft_PDS} (bottom panel) for the related power  spectra].
The model is   a  superposition of two broad Lorentzians (L$_b$, L$_h$) plus one or two 
narrow Lorentzians (L$_{QPO1}$, L$_{QPO2}$).
Parameters in square parentheses are frozen in the fits.
 Errors are given at the 90\% confidence level. See text about the identification of the $L_h$ component.}
\tablehead{\colhead{Components} & \colhead{ 0.1--2 keV} & \colhead{ 2--10 keV } }
\startdata 
Hard State  (TOO1) &       &               &        \\
L$_b$, $\nu_{b\ max}$ (Hz) & $0.59_{-0.13}^{+0.06}$  & $0.38_{-0.04}^{+0.08}$ \\
L$_h$, $\nu_{h\ max}$ (Hz)$^a$	 & $23\pm 2$  & $2.4_{-0.3}^{+3.1}$  \\   
VLNF, $\nu_{VLNF\ max}$ (Hz)			  &  \nodata & \nodata  \\
L$_{QPO1}$, $\nu_{QPO\ 0}$ (Hz)     & $1.55\pm 0.08$  & $1.53\pm 0.04$ \\
L$_{QPO2}$,  $\nu_{QPO\ 0}$ (Hz)          &  \nodata & $3.08\pm 0.07$ \\
\hline
Soft State  (TOO2)  &       &               &        \\
L$_b$, $\nu_{b\ max}$ (Hz) 			  & $36\pm3$  & [35]\\
L$_h$, $\nu_{h\ max}$ (Hz)			  &  \nodata   &  \nodata \\   
VLNF, $\nu_{VLNF\ max}$ (Hz) & ($5_{-3}^{+5})\times10^{-2}$&$(4.6_{-1.5}^{+1.1})\times10^{-4}$ \\

\enddata 
\tablenotetext{a}{See section 4.1, for details about the identification of the $L_h$ component in 
the 0.1--2 and 2--10 keV energy intervals}.

\label{tbl-lorentzian}
\end{deluxetable} 

%
%
\begin{deluxetable}{llllllll} 
\tablecolumns{6} 
\tablewidth{0pc}
\rotate 
\tabletypesize{\footnotesize}
\tablecaption{Energy dependence of PS component characteristics in Hard state (TOO-1) 
for the Lorentzian model.  Parameters in square parentheses are frozen in the fits.
Errors are given at the 90\% confidence level.} 
\tablehead{ 
\colhead{Components} & \colhead{0.1--2 keV}   & \colhead{2--4 keV}    & \colhead{4--10 keV} & \colhead{8--15 keV}   & \colhead{15--200 keV}
}
\startdata 
 
L$_b$, $\nu_{b,max}$ (Hz) & $0.59_{-0.13}^{+0.06}$ & $0.38\pm 0.06$ & $0.36_{-0.04}^{+0.08}$  & 
		$0.38_{-0.06}^{+0.08}$ & $3.3_{-0.9}^{+1.0}$ \\
L$_h$, $\nu_{h,max}$ (Hz)$^a$  & $23\pm 2$ & $2.1\pm 0.2$ & $2.5_{-0.5}^{+1.7}$ & 
		$3.5_{-0.8}^{+1.4}$ & $>50$ \\
L$_{QPO1}$, $\nu_{0}$ (Hz)  & \nodata & \nodata  & $1.54_{-0.06}^{+0.07}$ & 
		$[1.54]$ & \nodata \\
L$_{QPO2}$,  $\nu_{0}$ (Hz) &\nodata & \nodata &   $3.03_{-0.09}^{+0.10}$& 
	$[3.03]$  &  \nodata  \\

\enddata 
\tablenotetext{a}{See section 4.1 for details about the identification of the $L_h$ component in the 0.1--2,
2--4, 4--10, 8--15 and 15--200 keV intervals}.
\label{tbl-energy_depend_Lorentzian}
\end{deluxetable} 

\clearpage

%
%
\begin{deluxetable}{llllllll} 
\tablecolumns{7} 
\tablewidth{0pc}
\rotate 
\tabletypesize{\footnotesize}
\tablecaption{Best model fit parameters  used to fit the power spectra of TOO-1
in different energy bands. The model: {\sc bknpl} + {\sc lorentzian1} + 
{\sc lorentzian2}. 
  Parameters in square parentheses are frozen in the fits.
Errors are given at the 90\% confidence level.} 
\tablehead{ 
 \colhead{Parameter} & \colhead{0.1--2 keV}   & \colhead{2--4 keV}    & \colhead{4--10 keV} & 
\colhead{2--10 keV}    & \colhead{8--15 keV}   & \colhead{15--200 keV}
}
\startdata 

		 $\alpha_1$ & [0.] & [0.] & [0.] & [0.] & [0.] & [0.] \\
             $\alpha_2$ & $0.7\pm0.1$ & $1.20\pm0.02$ & $1.05\pm0.04$ & $1.098\pm0.013$ &
	$0.96\pm0.03$ & $0.49\pm0.10$ \\
   $\nu_{br}$ (Hz) & $0.28\pm 0.03$ & $0.345\pm 0.016$ & $0.36\pm 0.03$ & $0.35\pm 0.01$ &
	$0.37\pm 0.04$ & $0.4\pm 0.4$ \\
		 $\nu_{L1\ 0}$ (Hz) & $14^{-7}_{+3}$ & $1.49\pm0.09$ & $1.52\pm0.04$ & $1.53\pm0.03$ & 
	$1.54\pm0.04$ & $1.54\pm0.15$ \\
		 FWHM$_{L1}$ (Hz) & $30^{-14}_{+18}$ & $1.1\pm0.3$ & $0.8\pm0.3$ & $0.9\pm0.2$ & 
	$0.46\pm0.18$ & $1.1\pm1.4$ \\
		 $\nu_{L2\ 0}$ (Hz) & \nodata & [2$\nu_{L1}$] & [2$\nu_{L1}$] & [2$\nu_{L1}$] & 
	[2$\nu_{L1}$] & [2$\nu_{L1}$]  \\
		 FWHM$_{L2}$ (Hz) & \nodata & $1.6\pm0.8$ & $0.9\pm0.3$ & $0.95\pm0.28$ & 
	$1.1\pm0.5$ & $1\pm 1$ \\
		$\chi^2$/dof & 76.6/86 & 144.1/140 & 117.8/106 & 192.6/182 & 
 	95.9/68 & 8.6/13 \\

\enddata 
\label{tbl-4}
\end{deluxetable} 

\clearpage

%
%
\begin{deluxetable}{lll} 
\tablecolumns{3} 
\tablewidth{0pc} 
\tablecaption{Best fit parameters of the models used to fit the power spectra in 
two different time intervals (I101--I107, I109--I112) of TOO-1 in 4--10 keV . 
Model: {\sc bknpl} + {\sc lorentzian1} + 
{\sc lorentzian2}. Parameters in square parentheses are frozen in the fits.
Errors are given at the 90\% confidence level.} 
\tablehead{ 
\colhead{Parameter} & \colhead{I101--I107}   & \colhead{I109--I112}}
\startdata 
$\alpha_1$ & [0.] & [0.]  \\
		 $\alpha_2$ & $0.961\pm0.025$ & $1.23\pm0.12$ \\
		 $\nu_{br}$ (Hz) & $0.35\pm 0.03$ & $0.37\pm 0.07$ \\
		 $\nu_{L1}$ (Hz) & $1.47\pm0.03$ & $1.66\pm0.08$ \\
		 FWHM$_{L1}$ (Hz) & $0.63\pm0.25$ & $1.2\pm0.7$ \\
		 $\nu_{L2}$ (Hz) & [2$\nu_1$] & [2$\nu_1$]  \\
		 FWHM$_{L2}$ (Hz) & $0.5\pm0.2$ & $0.9\pm0.6$ \\
		 $\chi^2$/dof & 110.4/107 & 32.6/42 \\ 
\enddata 
\label{tbl-5}
\end{deluxetable}

\clearpage
%
%
\begin{deluxetable}{llll} 
\tablecolumns{4} 
\tablewidth{0pc} 
\rotate
\tablecaption{Best fit parameters of the models used to describe the power spectra of
TOO-2 (HSS). In 0.1-2 keV, two models: 1st model: {\sc pl} + {\sc bknpl}; 2nd model: 
PL + constant. The first index of the {\sc bknpl} was fixed to 0. 
In 2--10 keV. Errors are given at the 90\% confidence level.}
\tablehead{\colhead{Model} & \colhead{Parameter} & \colhead{0.1--2 keV}   & \colhead{2--10 keV}}
\startdata 
{\sc pl}$+${\sc bknpl} &       &               &        \\
			  & $\alpha$ & $1.0 \pm 0.3$ & \nodata \\
			  & $N_{PL}$  & $(4.0^{+6.0}_{-2.5})\times 10^{-4}$ & \nodata \\   
			  & $\alpha_1$ & [0.] & \nodata \\
			   & $\alpha_2$ & $1.0^{+0.1}_{-0.3}$ & \nodata \\
			   & $\nu_{br}$ (Hz) & $17^{+7}_{-2}$  & \nodata \\
			   & $N_{BKNPL}$ & $(2.0^{+0.1}_{-0.4})\times 10^{-3}$  & \nodata  \\
{\sc pl}$+${\sc const} &       &               &        \\
			   & $\alpha$ & \nodata & $2 \pm 1$ \\
			     & $N_{PL}$  & \nodata & $(3^{+100}_{-1})\times 10^{-8}$ \\
			     & $CONST$ & \nodata  & $(1.29 \pm 0.35)\times 10^{-5}$ \\   	   
			   & $\chi^2$/dof & 126/102 & 16.6/20 \\
\enddata 
\label{tbl-6}
\end{deluxetable}

\clearpage
%
%
\begin{figure}[ptbptbptb]
\includegraphics[scale=0.6, angle=-90]{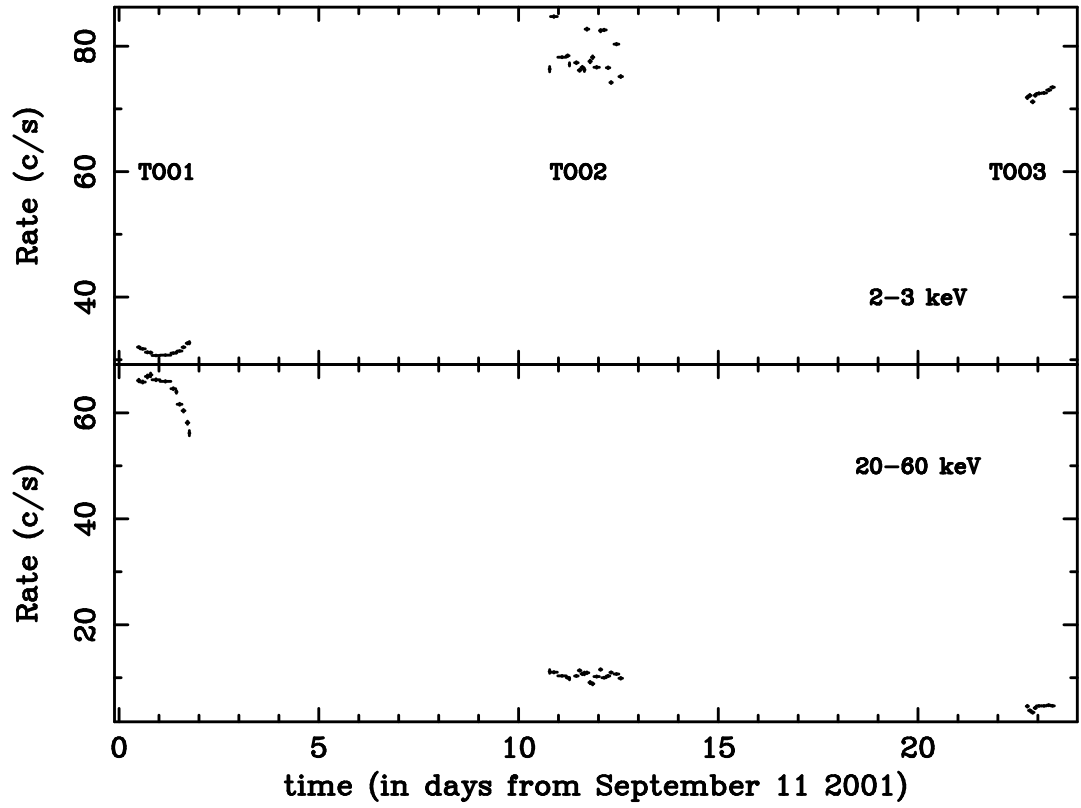}
\caption{Light curves  of XTE J1650-500 in two energy bands: 2-3 keV ({\em top}),   
20-60 keV ({\em bottom}).}
\label{lc}
\end{figure}

\clearpage
%
%
\begin{figure}[ptbptbptb]
\includegraphics[scale=0.6, angle=-90]{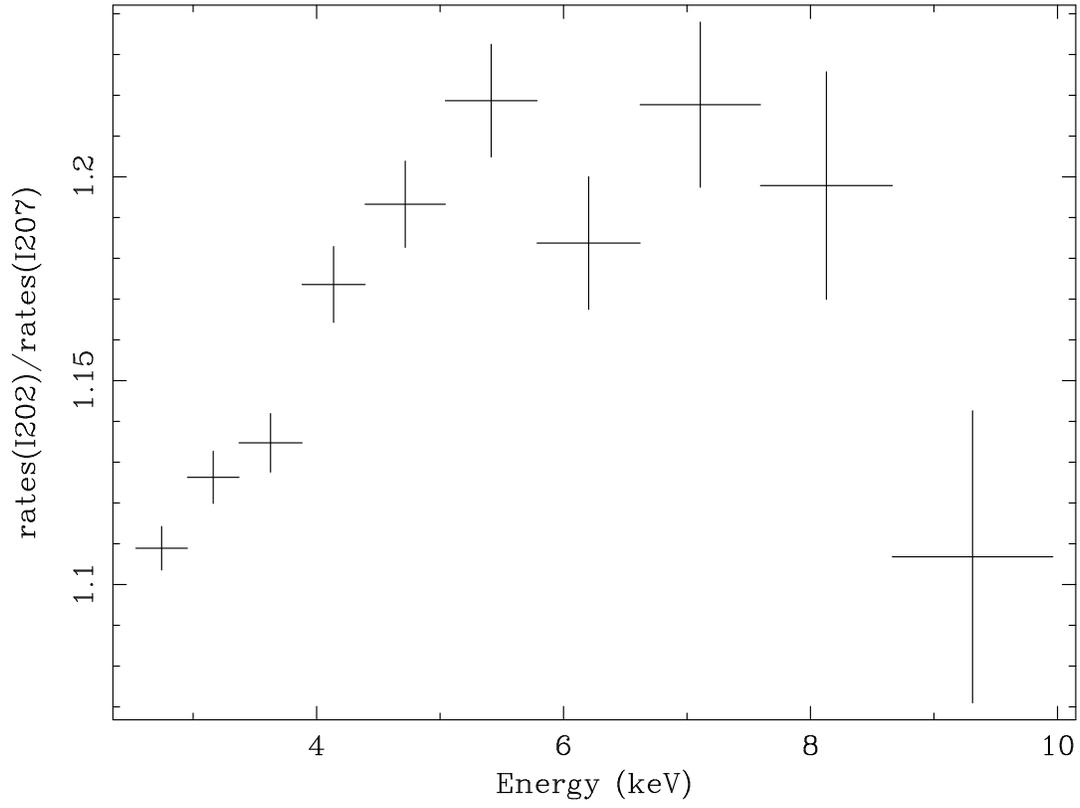}
\caption{Ratio between MECS mean count rate in the time interval I202 (TOO-2) and that in 
interval I207 (TOO-2)  as a function of energy.}
\label{rat}
\end{figure}

\clearpage
%
%
\begin{figure}[ptbptbptb]
\includegraphics[scale=0.6, angle=-90]{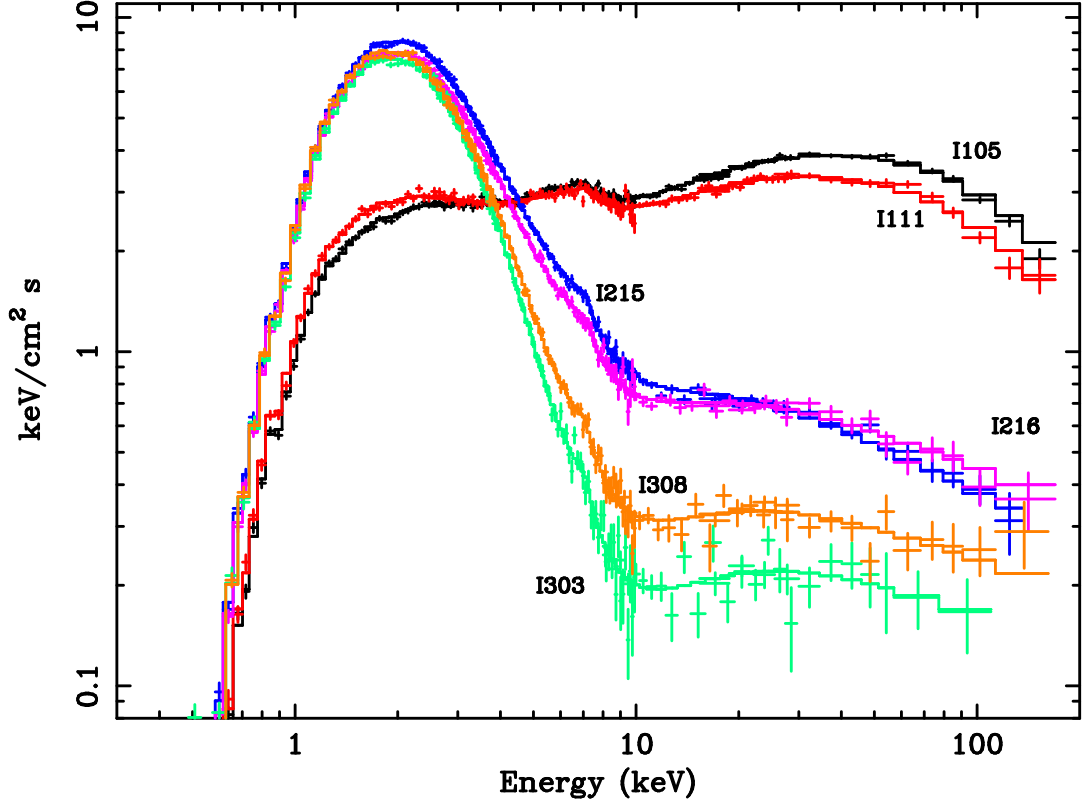}
\caption{Evolution of the  $EF(E)$ energy spectra with time. 
The  LHS spectra (I105 and I111) show the key role of the thermal Comptonization of 
soft (disk) photons in the extended hot Compton cloud (corona), while 
 the HSS spectra (I215--I216, I303, and I308) show that prominent  contribution
of the soft blackbody--like component due to the direct emission from the disk. 
In this case the hard X--ray component (i.e., the steep 
{\sc pl}) is formed in the relatively cold (temperature of a few keV) compact region 
which is presumably a converging flow into the BH. 
}
\label{sp_evol}
\end{figure}

\clearpage
%
%
\begin{figure}[ptbptbptb]
\includegraphics[scale=0.6, angle=-90]{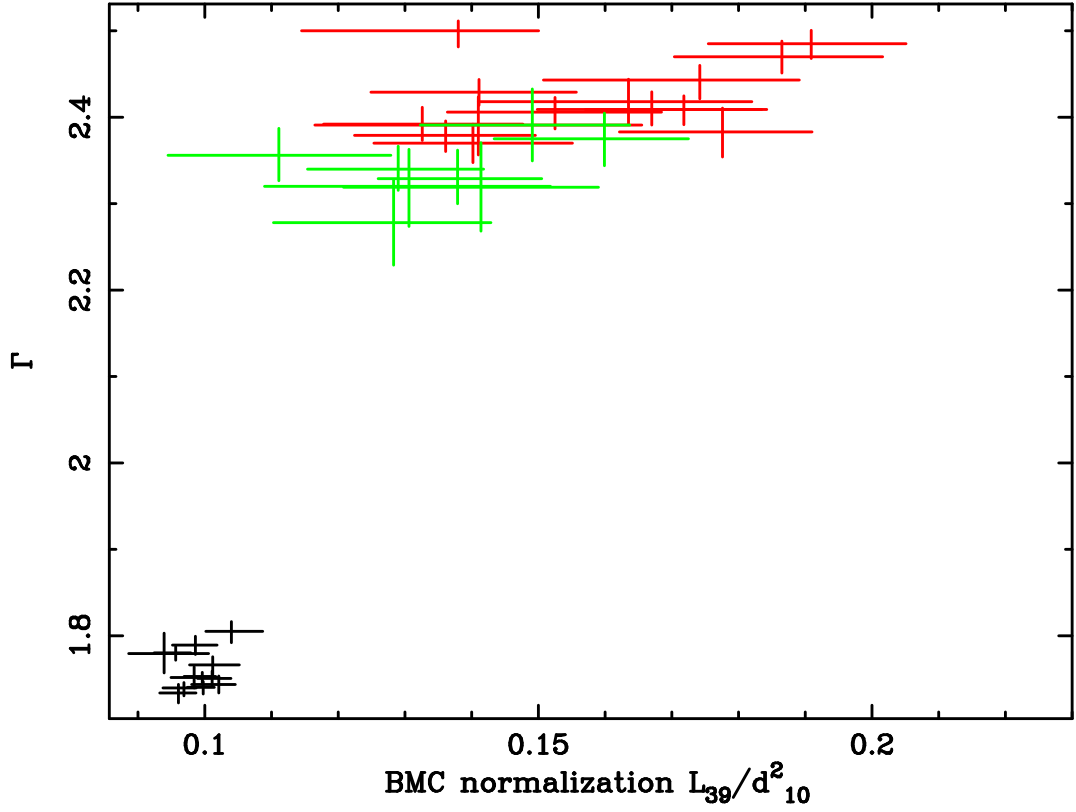}
\caption{Photon index evolution as a function of BMC normalization $L_{39}/d_{10}^2$ (which is proportional to   mass  accretion rate in the disk). Colors correspond to TOO1(black), TOO2 (red), TOO3 (green). 
  }
\label{index_vs_bmcnorm}
\end{figure}

\clearpage
%
%
\begin{figure}[ptbptbptb]
\includegraphics[scale=0.6, angle=-90]{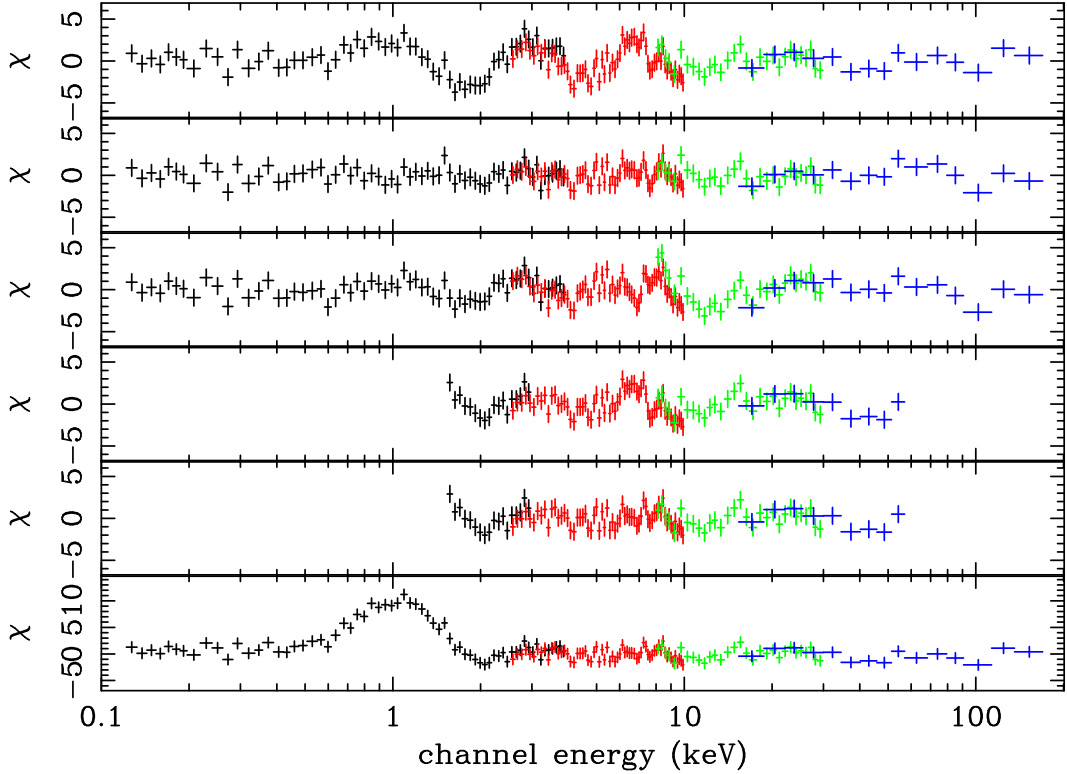}
\caption{Residuals (in standard deviation units) of the source spectrum in the time
interval I108 using different best-fit spectral models in different energy bands (see text).
{ From top to bottom. {\em Panel 1}: XSPEC model 
$WABS\times (BB+BMC)\times HIGHENCUT$ fits to data in the 0.1-200 keV range, 
 $\chi^2/$(dof)=430/165. 
{\em Panel 2}: our best-fit model ($WABS\times (BB+BMC)\times SMEDGE \times HIGHENCUT$)
fits to the same data, $\chi^2/$(dof)=150/164.
{\em Panel 3}: XSPEC model $WABS\times (BB+BMC+LAOR)\times HIGHENCUT$,
$\chi^2/$(dof)=300/163 fits to the data in the 0.1-200 keV.
{\em Panel 4}:  The same spectral
model, adopted to fit the spectrum of panel 1, is used to fit the spectrum in the 1.5-60 keV, 
$\chi^2/$(dof)=200/111. 
{\em Panel 5}: The same spectral
model, adopted to fit the spectrum of panel 3, is used to fit the spectrum in the 1.5-60 keV 
energy range, $\chi^2/$(dof)=136/109. 
{\em Panel 6}: The same model and parameters of panel 5, is used to fit the 0.1-200 keV 
energy range, $\chi^2/$(dof)=1315/164.}
}
\label{range}
\end{figure}

\clearpage
%
%
\begin{figure}[ptbptbptb]
\includegraphics[scale=0.35, angle=-90]{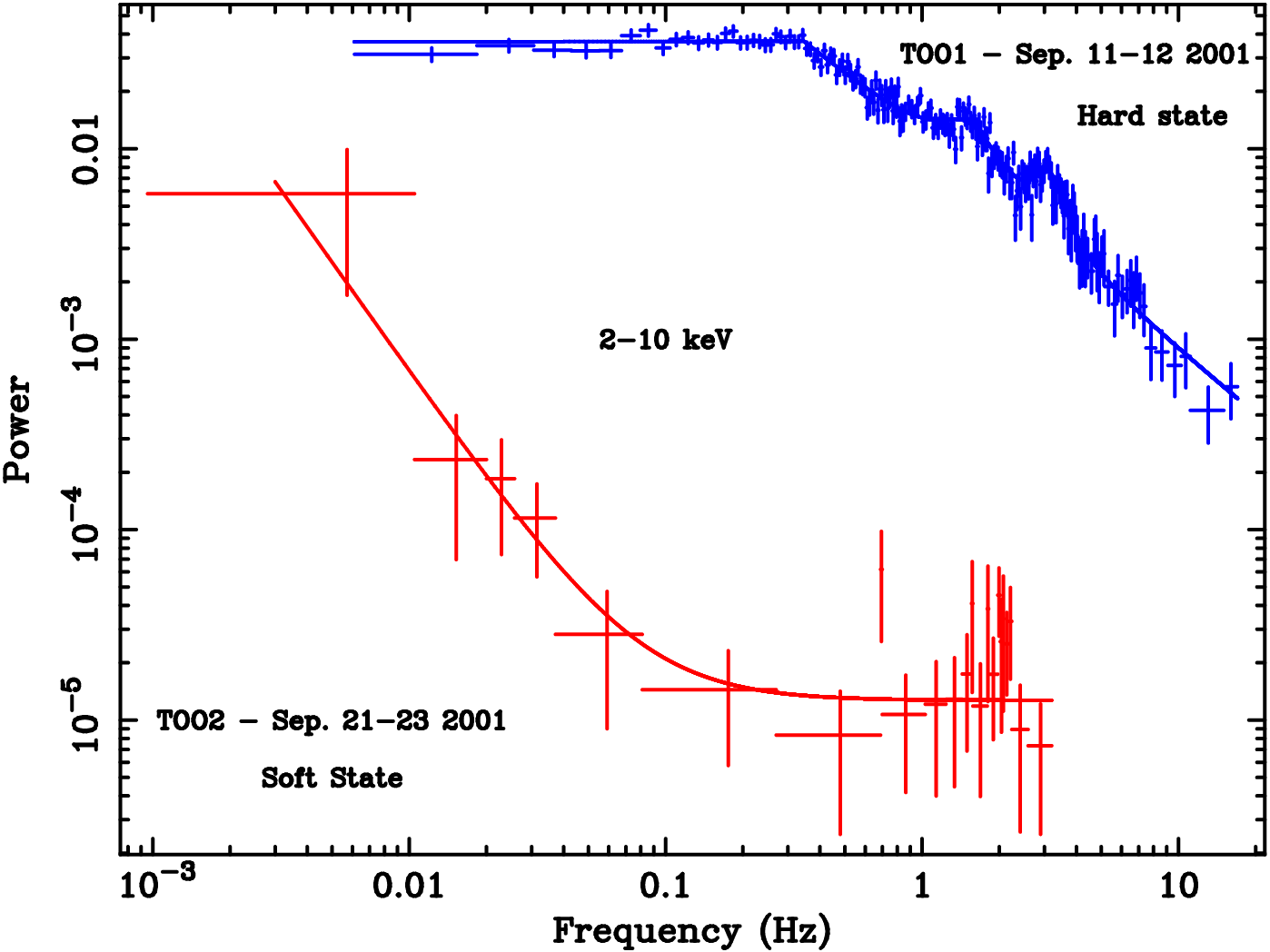}%
\includegraphics[scale=0.35, angle=-90]{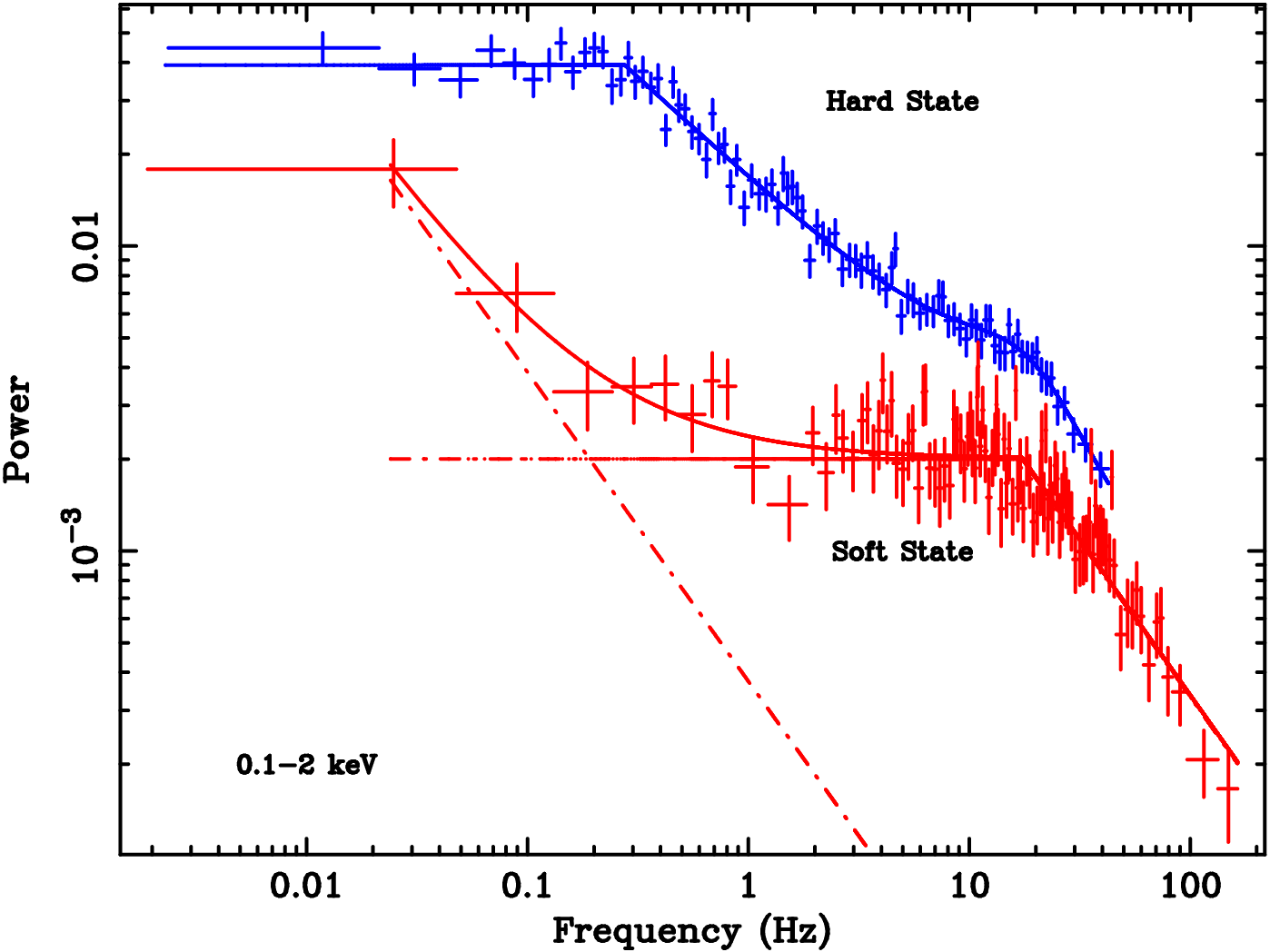}
\includegraphics[scale=0.35, angle=-90]{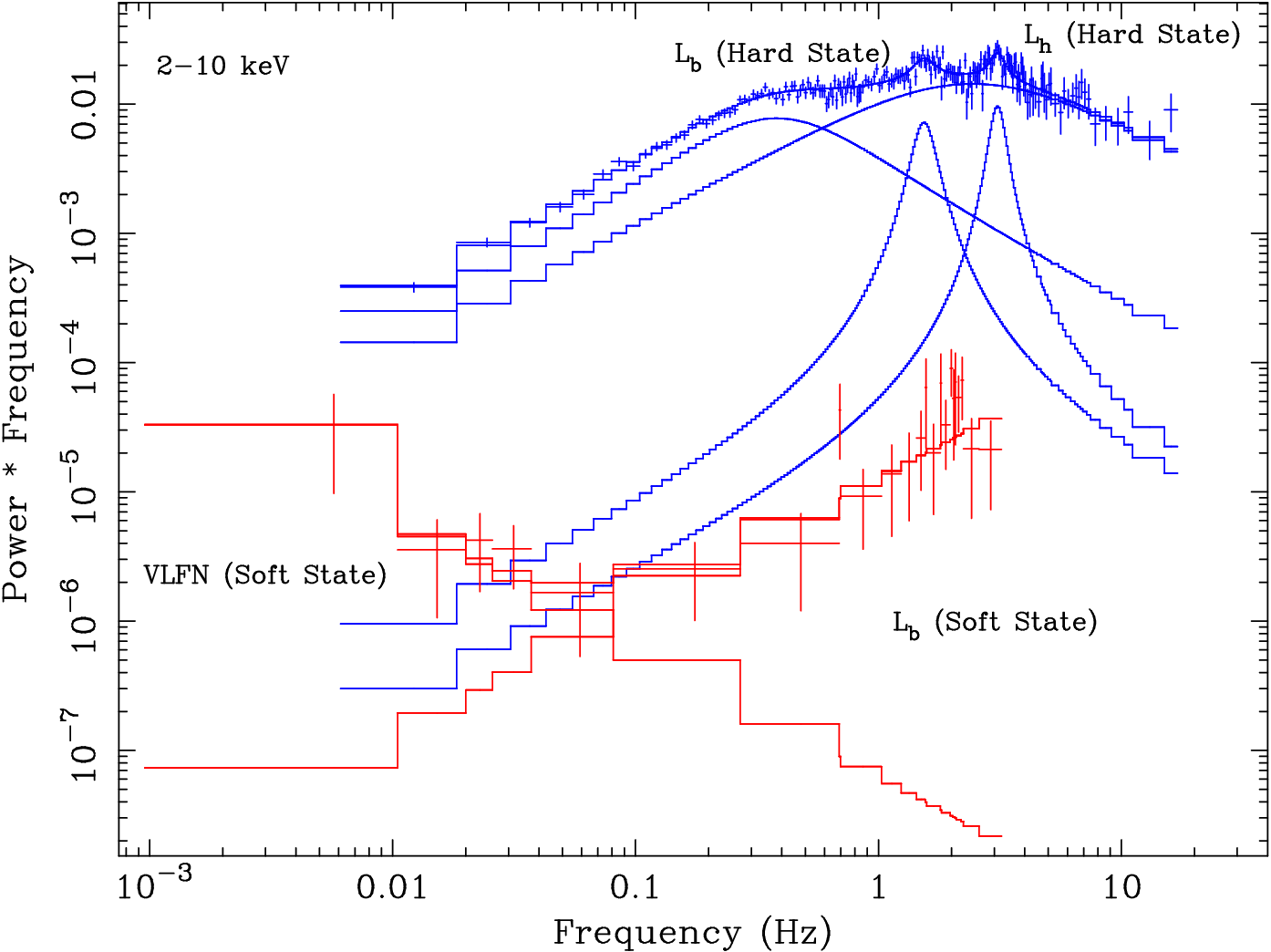}%
\includegraphics[scale=0.35, angle=-90]{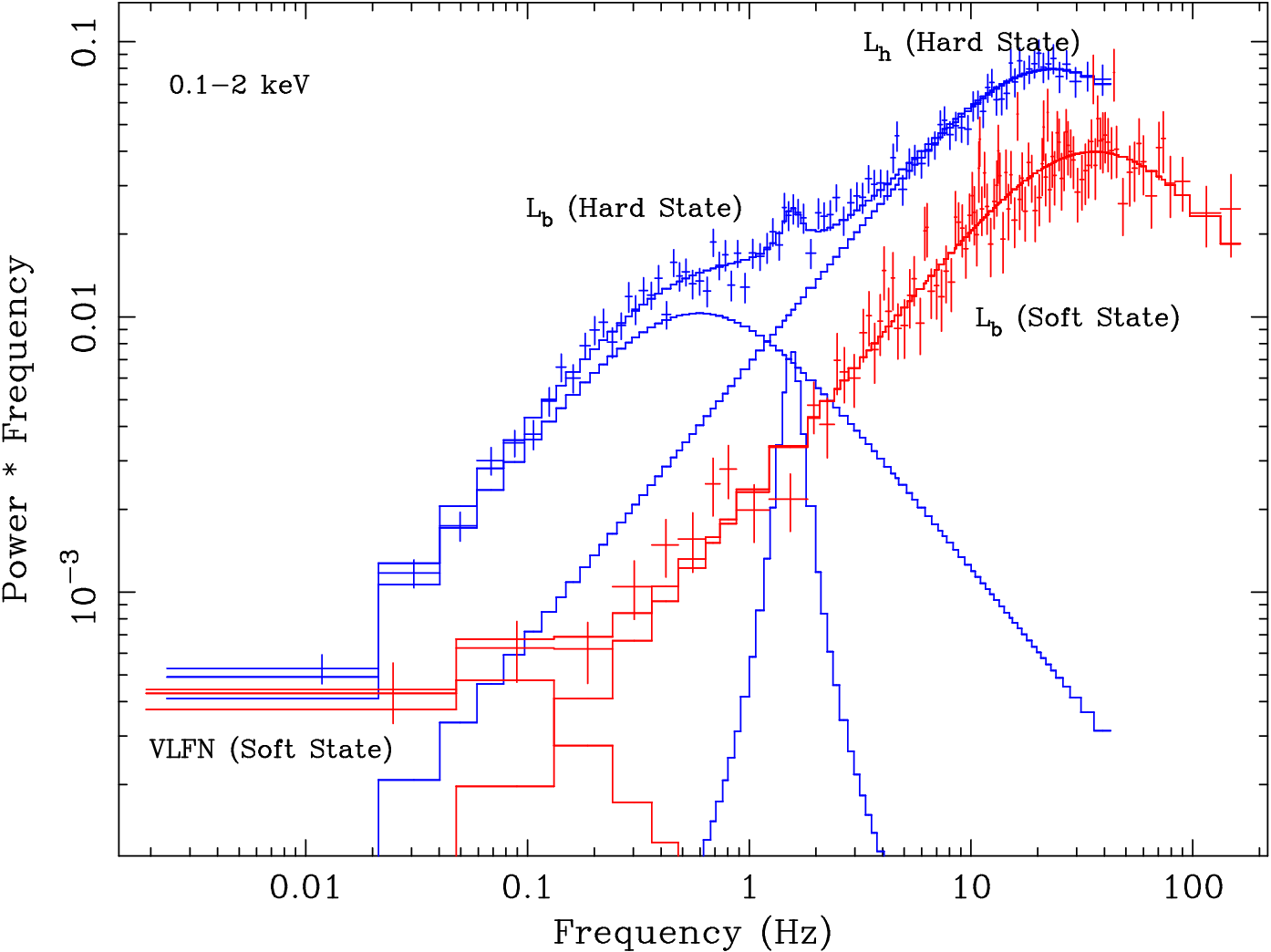}
\caption{The LHS PS (blue) vs. HSS PS (red) for two different energy 
bands: 2-10 keV  ({\em left panels}) and 0.1-2 keV ({\em right panels}). 
In the 2--10 keV,  {\it top panels}: the HSS PS is fitted by the sum of a {\sc pl} plus a 
{\sc constant} (see Table~\ref{tbl-6}), while the LHS PS is fitted by a {\sc bknpl} 
plus two Lorentzians (see Fig. 5); {\it bottom panel}, frequency$\times$ power diagram: the HSS PS is fitted by the sum of two broad
Lorentzians, while the LHS PS is fitted by the sum of 2 broad Lorentzians plus two narrow
Lorentzians.
It is apparent that the relative 
power of the HSS PS (with respect to the LHS PS) strongly depends on the energy band. 
}
\label{hard_vs_soft_PDS}
\end{figure}

\clearpage

%
%
\begin{figure}[ptbptbptb]
\includegraphics[scale=0.6,angle=-90]{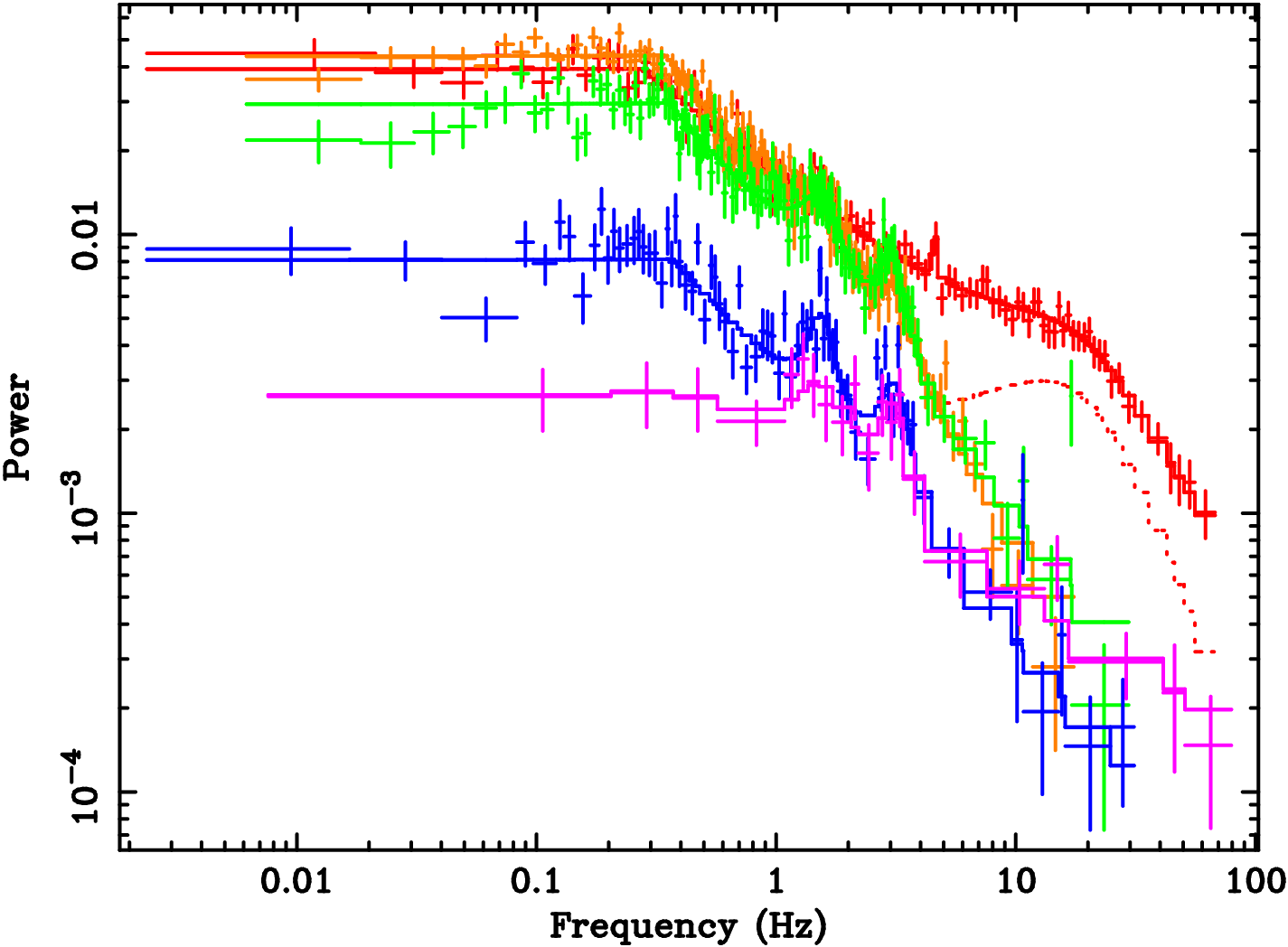}
\includegraphics[scale=0.6,angle=-90]{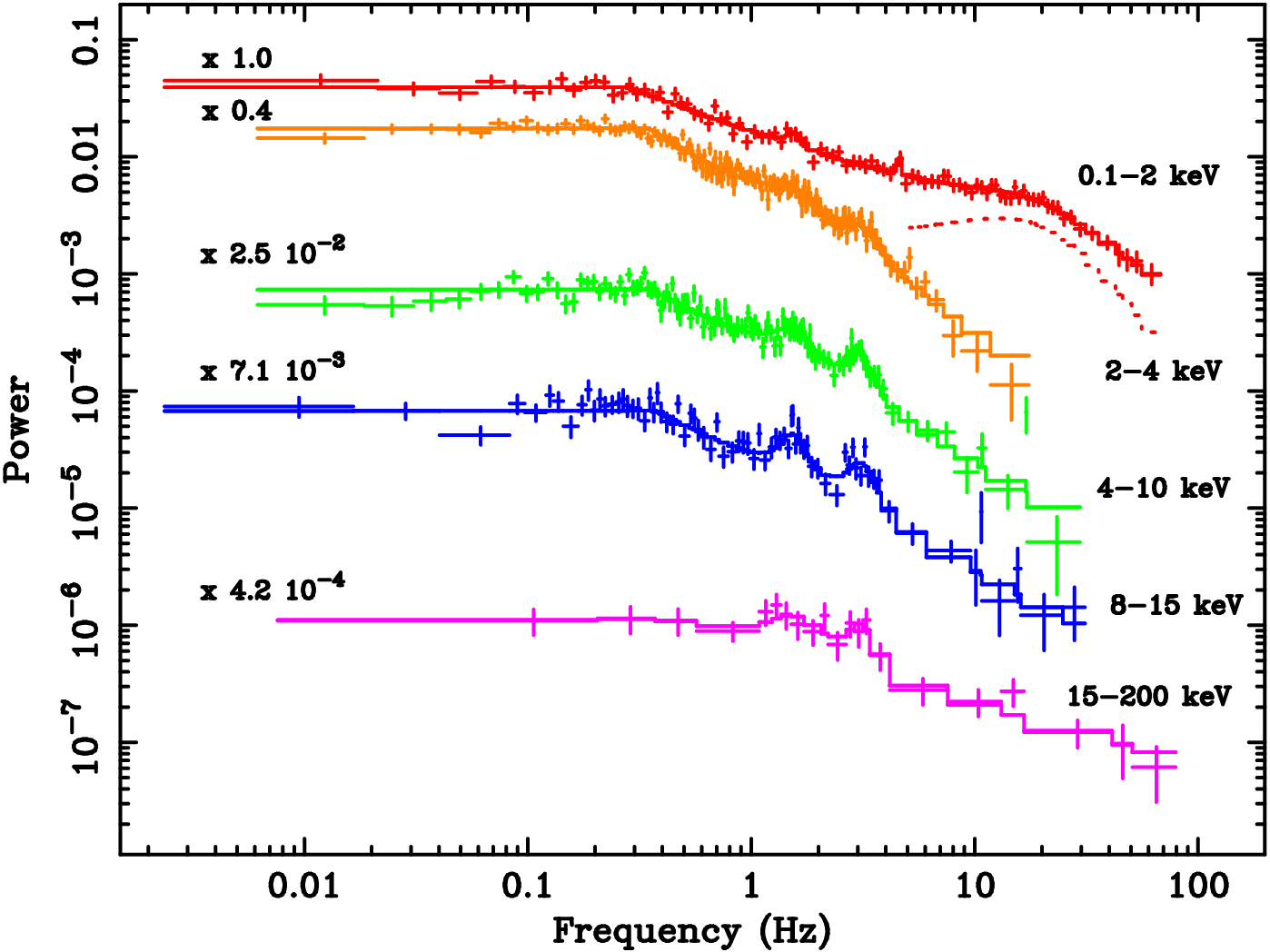}
\caption{ {\em Top panel}: the LHS power spectra in  different energy bands (see bottom panel for  the correspondence between  color  and energy intervals).  The decrease of the variability power 
with energy is evident from this Figure.
{\em Bottom panel}: the LHS power spectra as above, shifted, for clarity of display, 
along the ordinate 
direction by different factors, depending on the energy interval: factor 1 (0.1--2 keV), 
0.4 (2--4 keV), $2.5 \times 10^{-2}$ (4--10 keV), $7.1 \times10^{-3}$ (8--15 keV), and 
$4.2\times  10^{-4}$ (15--200 keV). 
} 
\label{pds_vs_energy}
\end{figure}

\clearpage
%
%
\begin{figure}[ptbptbptb]
\includegraphics[scale=0.6,angle=-90]{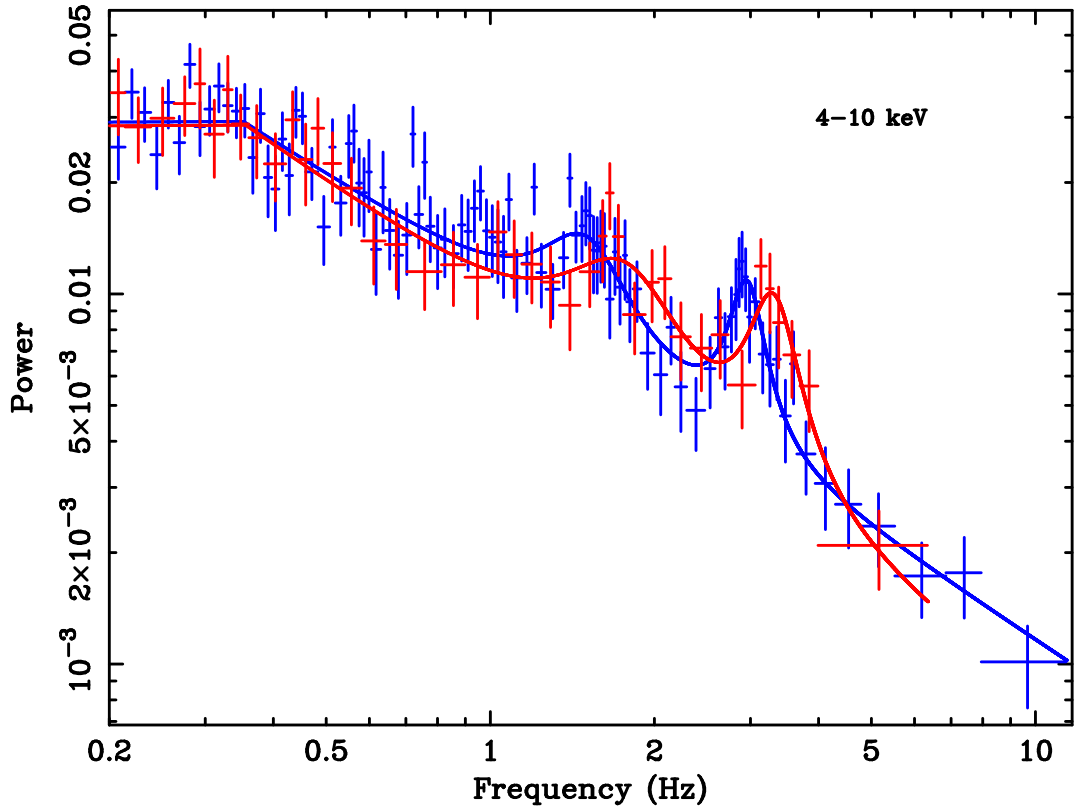}
\caption{The LHS power spectra  of XTE J1650-500 in the time intervals
from I101 to I107 and from I109 to I112 (see text for details). 
The power spectrum (red line) corresponding to the softer energy spectrum is shifted 
to higher frequencies with respect to that (blue line) corresponding to the hard energy 
spectrum (see related energy spectra in Fig. \ref{sp_evol}). 
The fit with 
a {\sc bknpl} plus two Lorentzians  is also shown. Best fit details in Table~\ref{tbl-4}.
}
\label{pds_shift}
\end{figure}

\clearpage
%
%
\begin{figure}[ptbptbptb]
\includegraphics[scale=0.6, angle=-90]{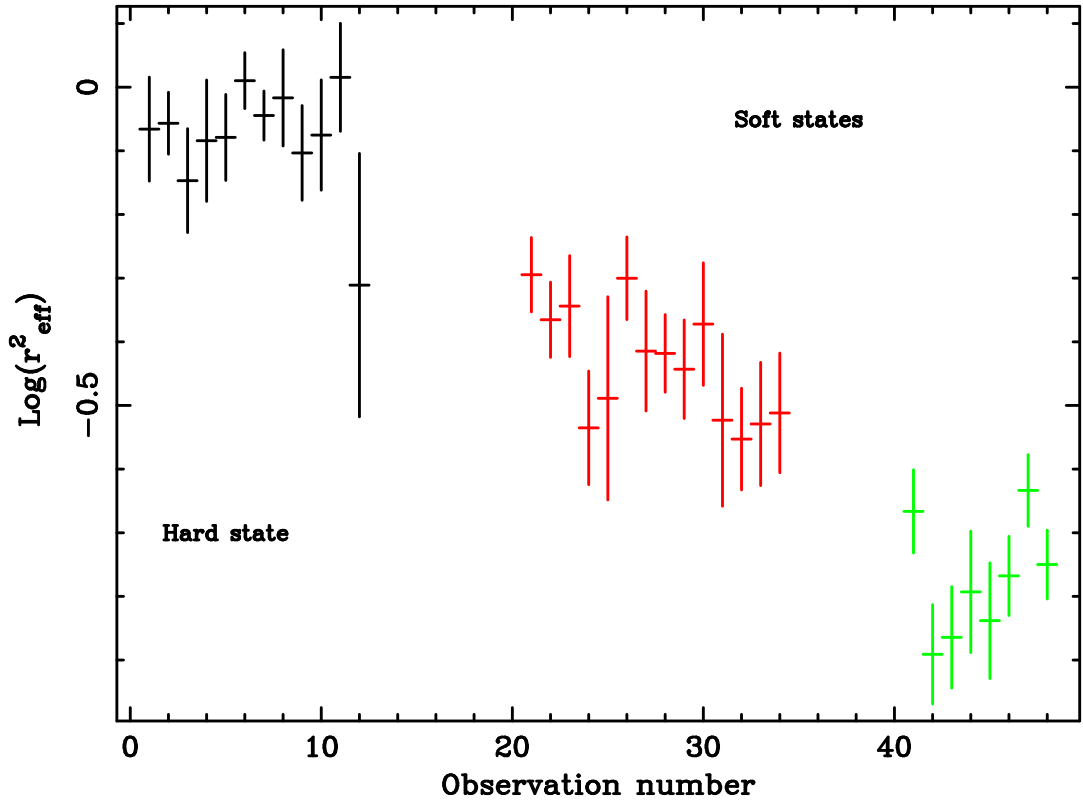}
\caption{Evolution of the Compton cloud area during the spectral transition exemplified 
by the progressive number of the time intervals from TOO--1 to TOO--3 (see text). Colors correspond to TOO1 (black), TOO2 (red), TOO3 (green). 
  }
\label{sp_evol_area}
\end{figure}

\clearpage
%
%
\begin{figure}[ptbptbptb]
\includegraphics[scale=0.6, angle=-90]{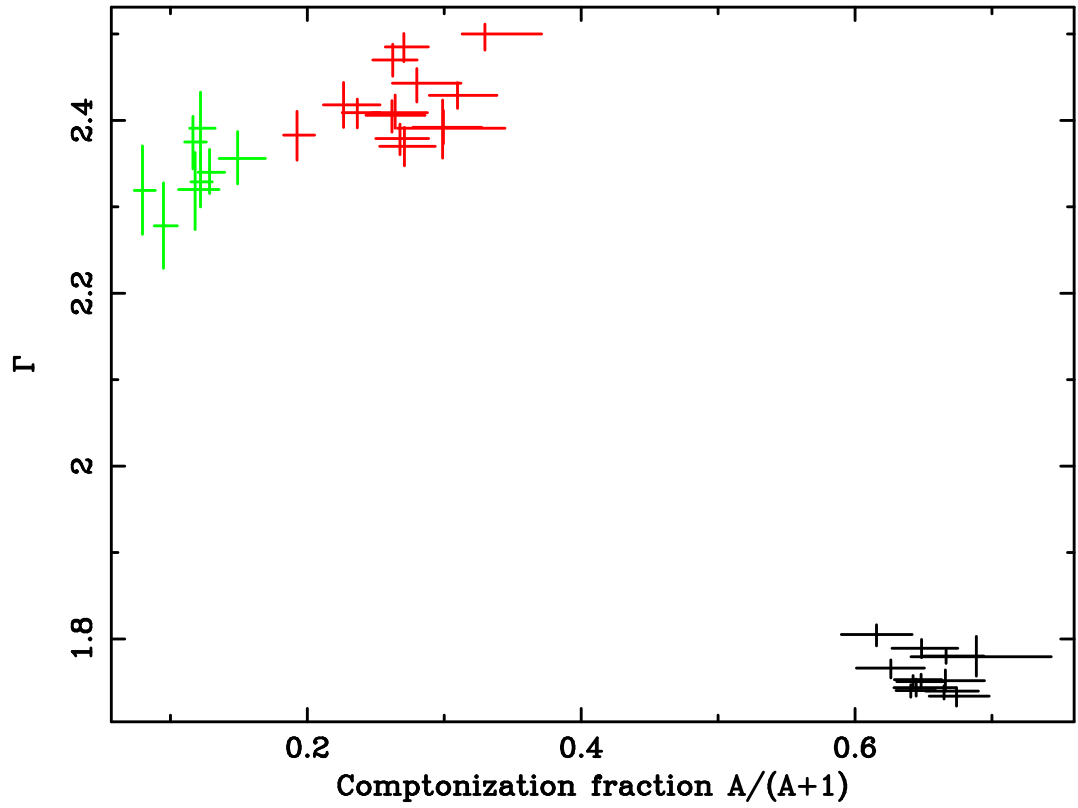}
\caption{Evolution of the Comptonization fraction $f=A/(1+A)$ during the spectral transition from LHS  to HSS. Colors correspond to TOO1 (black), TOO2 (red), TOO3 (green). 
  }
\label{comp_fractevol_vs_index}
\end{figure}


\begin{thebibliography}{}

\bibitem[Arnaud (1996)]{Arnaud96} Arnaud, K.~A. 1996, ASP Conf. Series, 101, 17

\bibitem[Basko \etal (1974)]{bst74} Basko, M. M., Sunyaev, R.A. \& Titarchuk, L.G.  
1974, A\&A, 31, 249
\bibitem[Basko (1978)]{basko} Basko, M. M. 1978, \apj, 223, 268


\bibitem[Belloni \etal (2002)]{bpk02} 
 Belloni, T., Psaltis, D.,  \&  van der Klis, M 2002, \apj, 572, 392

\bibitem[Boella \etal (1997a)]{boella} Boella, G., Butler, R. C.,
  Perola, G. C., Piro, L., Scarsi, L., Bleeker, J. A. M., 1997a, \aaps, 122, 299

\bibitem[Boella \etal (1997b)]{boellab} Boella, G., Chiappetti, L.,
  Conti, G., Cusumano, G., del Sordo, S., La Rosa, G., Maccarone,
  M. C., Mineo, T., Molendi, S., Re, S., Sacco, B., Tripiciano, M.,
  1997b, \aaps, 122, 327

\bibitem[Borozdin \etal (1999)]{BRT99}
Borozdin, K., Revnivtsev, M.,  Trudolyubov, S., Shrader, C. \& Titarchuk, L. 1999, 
\apj, 517, 367 

\bibitem[Bradshaw \etal (2007)]{btk07}
Bradshaw, C. F.,  Titarchuk, L., \&  Kuznetsov, S. 2007, \apj, 663, 1225  


\bibitem[Corbel \etal (2004)]{corb04}
Corbel, S., Fender, R. P., Tomsick, J. A., Tzioumis, A. K., \& Tingay, S. 2004,
ApJ, 617, 1272


\bibitem[Frontera \etal (1997)]{ff} Frontera, F., Costa, E., dal Fiume,
  D., Feroci, M., Nicastro, L., Orlandini, M., Palazzi, E., Zavattini,
  G., 1997, \aaps, 122, 357


\bibitem[Gies \& Bolton (1982)]{gb82} Gies, D. R.,  \& Bolton, C. T. 1982, \apj, 260, 240




\bibitem[Groot et al. (2001)]{groot01}
Groot, P., Tingay, S., Udalski, A., \& Miller, J. 2001, IAU Circ., 7708, 4


\bibitem[Grove et al. (1998)]{grove} Grove, J. E., Johnson, W.N.,  Kroeger, R. A.,  McNaron-Brown, K. \& 
Skibo, J.G.  1998, \apj, 500, 899

\bibitem[Homan \etal (2003)]{h03}
Homan, J. et al.  2003, ApJ, 586, 1262

\bibitem[Hua \& Titarchuk  (1995)]{ht95}  
Hua, X-M., \&  Titarchuk, L.  1995,  \apj, 449, 188 

\bibitem[Kalemci et al. (2003)]{k03}  
Kalemci, M. et al.    2003,  \apj, 586, 419 

 








  


 


\bibitem[Kallman et al. (2004)]{k04}Kallman, T.R., Palmeri, P.,  Bautista, M.A., 
Mendoza, C.  \& Krolik, J.H. 2004, ApJS, 155, 675

\bibitem[Klein-Wolt  \& van der Klis (2008)]{kvk08}
Klein-Wolt, M.  \& van der Klis, M. 2008, ApJ, 675, 1407 (KVK08)

\bibitem[Laming \& Titarchuk (2004)]{lt04}
Laming, J.M. \& Titarchuk, L.   2004,  \apj, 615, 121  

\bibitem[Laurent \& Titarchuk (2007)]{lt07}
Laurent, P. \& Titarchuk, L.   2007,  \apj, 656, 1056  

\bibitem[Leahy et al. (1983)]{lea83} Leahy, D. A., Darbro, W., Elsner, R. F. 
\etal, 1983, \apj, 266, 160

\bibitem[Lyubarskii (1997)]{L97}
Lyubarskii, Yu., E.  1997,  MNRAS, 292, 679 


\bibitem[Manzo \etal (1997)]{manzo} Manzo, G., Giarrusso, S.,
  Santangelo, A., Ciralli, F., Fazio, G., Piraino, S., \& Segreto, A.
  1997, \aaps, 122, 341

\bibitem[Markwardt et al.  (2001)]{mark01}
Markwardt, C., Swank, J., \& Smith, E. 2001, IAU Circ., 7707, 2






 



\bibitem[Miller \etal (2002)]{mil02} Miller, J. M., Fabian, A. C.,
Wijnands, R., Reynolds, C. S., Ehle, M., Freyberg, M. J., van der Klis, M.,
Lewin, W., H., G., Sanchez--Fernandez, C., \& Castro--Tirado, A. J. 2002,
\apj, 570, L69




\bibitem[Miller et al. (2004)]{mil04} Miller, J.~M.  et al. 2004, \apj,  606, L131

\bibitem[Miniutti \etal (2004)]{min} Miniutti, G., Fabian, A. C.,  \& Miller, J. M. 2004, 
MNRAS, 351, 466 (M04)

\bibitem[Montanari \etal (2004)]{mon} Montanari, E., Frontera, F., \&
Amati, L. 2004, Nucl. Phys. B Proc. Suppl., 132, 412







\bibitem[Orosz \etal (2004)]{oro04} Orosz, J. A., McClintock, J. E., Remillard,  
R. A., \& Corbel, S., 2004, \apj, 616, 376

\bibitem[Parmar \etal (1997)]{parmar} Parmar, A. N., Martin, D. D. E.,
  Bavdaz, M., Favata, F., Kuulkers, E., Vacanti, G., Lammers, U.,
  Peacock, A., \& Taylor, B. G., 1997, \aaps, 122, 309

\bibitem[Psaltis \etal (1999)]{pbk99} 
Psaltis, D.,  Belloni, T., \&  van der Klis, M 1999, \apj, 520, 262



\bibitem [Remillard et al. (2001)]{rem01} 
Remillard, R. A. 2001, IAU Circ., 7707, 1

\bibitem [Revnivtsev \& Sunyaev  (2001)]{rs01} 
Revnivtsev, M., \& Sunyaev, R. 2001, IAU Circ., 7715, 1

\bibitem[Rossi \etal (2005)]{ros05} Rossi, S., Homan, J., Miller, J. M., \&
Belloni, T. 2005, \mnras, 360, 763

\bibitem[Rossi \etal (2004)]{r04} Rossi, S., Homan, J., Miller, J. M., \&
Belloni, T. 2004, NuPhS, 132, 416

\bibitem[Shakura \& Sunyaev  (1973)]{ss73} Shakura, N.I., \& Sunyaev, R.A. 1973, \aap, 24, 337

\bibitem[Shaposhnikov \& Titarchuk (2007)] {st07} Shaposhnikov, N., \& Titarchuk, L. 2007, 
\apj, 663, 445 (ST07)

\bibitem[Shaposhnikov \& Titarchuk (2006)]{ST06} Shaposhnikov, N., \& Titarchuk, L. 2006, 
\apj, 643, 1098 (ST06)

\bibitem[Shaposhnikov \& Titarchuk (2008)]{st08}
Shaposhnikov, N., \& Titarchuk, L. 2008,  submitted to \apj.


\bibitem[Shrader \& Titarchuk (1999)]{st99}
 Shrader, C., \& Titarchuk, L.G. 1999,  ApJ, 521, L121 

 














\bibitem[Titarchuk (1994)]{t94}
Titarchuk, L.  1994,  \apj, 434, 272

\bibitem[Titarchuk  \etal  (2007)]{tks07}
Titarchuk, L., Kuznetsov, S. \& Shaposhnikov, N.  2007,  \apj, 667, 

\bibitem[Titarchuk, \& Fiorito (2004)]{TF04}
Titarchuk, L.G. \& Fiorito, R.  2004,  \apj, 612,  988 (TF04)  

\bibitem[Titarchuk \etal (1998)]{TLM98}
Titarchuk, L., Lapidus, I.I., \& Muslimov, A. 1998, \apj,  499, 315 (TLM98)

\bibitem[Titarchuk \& Lyubarskij  (1995)]{tl95}  
Titarchuk, L., \& Lyubarskij, Yu.  1995,  \apj, 450, 876 

\bibitem[Titarchuk \etal (1997)]{bmc}  Titarchuk, L., Mastichiadis, A., \& Kylafis, N. D., 1997,  \apj, 487, 834


\bibitem[Titarchuk, Mastichiadis \&  Kylafis, (1996)]{tmk96}
 Titarchuk, L. G., Mastichiadis, A., \& Kylafis, N. D. 1996, \aap, 120, 171 (TMK96)
 







\bibitem[Titarchuk \& Zannias (1998)]{TZ98} Titarchuk, L.G. \& Zannias, T. 
1998, 493, 863




\bibitem[Titarchuk \& Shaposhnikov (2008)]{ts08}
Titarchuk, L.G. \& Shaposhnikov, N.  2008,  \apj,   678,  1230 (TS08)

\bibitem[Titarchuk \etal  (2007)]{tsa07}
Titarchuk, L.G., Shaposhnikov, N. \& Arefiev, V. 2007,  \apj,  660, 556 (TSA07)

\bibitem[Titarchuk, \& Shaposhnikov (2005)]{ts05}
Titarchuk, L.G. \& Shaposhnikov, N.  2005,  \apj,  626, 298 (TS05)

\bibitem[Uttley, McHardy \& Vaughan (2005)]{utt05} 
Uttley, P.,  McHardy, I.M.,   \& Vaughan, S.  2005, MNRAS, 359, 345

\bibitem[van der Klis (1989)]{klis} van der Klis, M. 1989, Timing neutron
stars, ed. H. \"Ogelman, \& E.P.J. van der Heuvel (Dordrech:Kluwer), NATO ASI
Ser. C, 262, 27 
 
\bibitem[van der Klis (1995)]{klis95} van der Klis, M.  1995, in X-ray binaries, 
ed. W.H.G. Lewin, J.,van Paradijs, \& E.P.J. van der Heuvel (Cambridge Univ.  Press, 
Cambridge), p. 252

\bibitem[Vignarca et al. (2003)]{vig}
Vignarca, F., Migliari, S., Belloni, T., Psaltis, D., \& van der Klis, M. 2003,  A\&A, 
397, 729 (V03)


\bibitem[Wijnands et al. (2001)]{w01}
Wijnands, R., Miller, J. M., \& Lewin, W. H. G. 2001, IAU Circ., 7715, 2

\bibitem[Wijnands \& van der Klis  (1999)]{wk99}
Wijnands, R.,  van der Klis, M.  1999, \apj,  514, 939

\bibitem[Wilms et al.  (2006)]{w06}
Wilms, J., Nowak, M. A., Pottschmidt, K., Pooley, G. G., \& Fritz, S.  2006, A\&A, 447, 245





\end{thebibliography}
\end{document}